\documentclass[journal = ancac3, layout=twocolumn]{achemso}

\usepackage[english]{babel}
\usepackage{graphicx}
\usepackage{siunitx}
\usepackage{amsmath}
\usepackage{amssymb}
\usepackage[]{url}

\newcommand{\laserlaser}{\textit{laser$\times$laser}}
\newcommand{\laserspp}{\textit{laser$\times$SPP}}
\newcommand{\sppspp}{\textit{SPP$\times$SPP}}

\newcommand{\titlestr}{Energy and Momentum Distribution of Surface Plasmon-induced Hot Carriers Isolated via Spatiotemporal Separation}

\author{Michael Hartelt} 
\email{hartelt@physik.uni-kl.de}

\author{Pavel N. Terekhin} 
\author{Tobias Eul} 
\author{Anna-Katharina Mahro} 
\author{Benjamin Frisch}
\author{Eva Prinz} 
\affiliation{Department of Physics and Research Center OPTIMAS, TU Kaiserslautern, Germany}

\author{Baerbel Rethfeld}
\affiliation{Department of Physics and Research Center OPTIMAS, TU Kaiserslautern, Germany}	

\author{Benjamin Stadtm\"uller} 
\affiliation{Department of Physics and Research Center OPTIMAS, TU Kaiserslautern, Germany}	

\author{Martin Aeschlimann} 
\affiliation{Department of Physics and Research Center OPTIMAS, TU Kaiserslautern, Germany}	
\title{\titlestr}
\keywords{plasmon-induced hot carriers; surface plasmon polariton; PEEM; 2-photon photoemission; momentum microscopy}

\begin{document} 


\date{\today}

\begin{abstract}
Understanding the differences 
between photon-induced and plasmon-induced hot electrons 
is essential for the construction of devices for plasmonic energy conversion.
The mechanism of the plasmonic enhancement 
in photochemistry, 
photocatalysis, 
and light-harvesting 
and especially the role of hot carriers is still heavily discussed. 
The question remains, 
if plasmon-induced and photon-induced hot carriers
are fundamentally different,
or if plasmonic enhancement 
is only an effect of field concentration
producing these carriers
in greater numbers.
For the bulk plasmon resonance,
a fundamental difference is known,
yet for the technologically important surface plasmons 
this is far from being settled.
The direct imaging of surface plasmon-induced hot carriers 
could provide essential insight,
but the separation of
the influence of 
driving laser, 
field-enhancement, 
and fundamental plasmon decay
has proven to be difficult.
Here, we present an approach
using a two-color femtosecond pump-probe scheme 
in time-resolved 2-photon-photoemission (tr-2PPE), 
supported by a theoretical analysis 
of the light and plasmon energy flow. 
We separate the energy and momentum distribution 
of the plasmon-induced hot electrons 
from the one of photoexcited electrons 
by following the spatial evolution of photoemitted electrons 
with energy-resolved Photoemission Electron Microscopy (PEEM) 
and Momentum Microscopy 
during the propagation of a Surface Plasmon Polariton (SPP) pulse along a gold surface. 
With this scheme, 
we realize a direct 
experimental access 
to plasmon-induced hot electrons.
We find a plasmonic enhancement 
towards high excitation energies
and small in-plane momenta,
which suggests 
a fundamentally different mechanism
of hot electron generation,
as previously unknown
for surface plasmons.
\end{abstract}


\maketitle

\section{Introduction}
The world's vital demand for clean energy  
poses a huge challenge for fundamental and application-oriented research 
to devise new and sustainable concepts 
to convert solar power into electric and chemical energy. 
One key perspective 
to enhance the efficiency 
of light-to-carrier conversion processes 
is plasmon technology. 
Despite the great potential of plasmon-induced hot carriers 
for photovoltaics and photochemistry, 
\cite{Brongersma_2015_NatureNanotechnology, 
	Linic_2011_NatureMaterials, 
	Atwater_2010_NatureMaterials}
their precise role 
in the enhancement 
of the efficiency 
of these processes is still heavily discussed.
\cite{Brongersma_2015_NatureNanotechnology, faradaydiscussions-theoryhotelectrons, Baumberg_2019_FaradayDiscussions}
Fundamental questions remain on 
how 
plasmon-induced hot carriers 
are generated, 
how they dissipate energy and momentum, 
and how the underlying mechanisms 
come into play 
in plasmonic energy conversion processes. 
Many theoretical studies have been conducted 
in the last years
to address these questions.
\cite{Manjavacas_2014_ACSNano, Sundararaman2014, Zhang2014, Bernardi_2015_Naturecommunications, brown2016acsnano, Saavedra_2016_ACSPhotonics, Sykes2017, Khurgin2019faradaydiscussions, Aguirregabiria_2019_FaradayDiscussions, Do_2021_Nanoscale, Benhayoun_2021_AppliedSurfaceScience, Khurgin_2021_ACSPhotonics}
However, 
it is essentially
unresolved 
if plasmonic enhancement 
is simply due to the field-enhancement 
of light at the surface of a metal,
or if there is a more fundamental difference
between plasmon-induced and photon-induced hot carriers.

For the bulk plasmon resonance 
at the plasma frequency $\omega_\mathrm{p}$,
\bibnote{
	In this frequency range, 
	the real part of the dielectric function passes zero
	(\textit{epsilon near zero}).}
a fundamental difference
has long been known from theoretical studies 
\cite{Hopfield_1965_PhysicalReview}
and was more recently shown in 
linear \cite{Barman_2004_SurfaceScience, Barman_2004_PhysicalReviewB}
and 
non-linear \cite{Reutzel_2019_PhysicalReviewLetters, Li_2021_ACSPhotonics}
photoemission experiments:
Bulk plasmons selectively excite electrons 
from occupied states 
close to the Fermi energy $E_\mathrm{F}$ of the metal.
This results in a peak in the photoemission spectrum at 
$E-E_\mathrm{F} = \hbar \omega_\mathrm{p}$ 
(or $2 \hbar \omega_\mathrm{p}$ in the non-linear case).
This energy-selective emission 
strongly contradicts 
the expectation for the electron and hole distributions 
in conventional photoexcitation, 
which are governed by the density of states (DOS) of the material. 
The effect was referred to as
\textit{non-Einsteinian} photoemission
\cite{Reutzel_2019_PhysicalReviewLetters, Li_2021_ACSPhotonics, Novko_2021_PhysicalReviewB}
due to its fixed energetic position
for all 
photon energies.

In energy harvesting applications, 
often energy thresholds have to be overcome 
for effective charge separation or driving a chemical reaction.
Therefore, 
a preferential high-energetic electron excitation,
as observed for bulk plasmons,
could hold great potential.
This would require 
that a similar effect 
also exists for surface plasmons,
which can be excited at optical frequencies
and 
concentrate the energy density at the metal surface,
where electrons can be 
transferred across a functional interface
or they can be excited directly 
\latin{via} 
chemical interface damping.
\cite{Kale_2015_Science, Khurgin_2021_ACSPhotonics}
A theoretical model 
for the bulk case 
by Novko 
\latin{et al.}
suggests
that a similar plasmonic decay
for surface plasmons 
might be possible.
\cite{Novko_2021_PhysicalReviewB}
Although 
some experiments at nanostructure interfaces
show an enhancement for high-energy electrons,
\cite{Tan_2017_NaturePhotonics, Tan_2018_PhysicalReviewLetters, Shibuta_2021_ACSNano}
a clear identification 
of the microscopic mechanisms 
was not yet possible 
due to the complex interplay 
of various experimental parameters 
(field enhancement, interface effects, laser field, 
\latin{etc.}) 

Conventionally, 
plasmonic excitations and their decay 
are 
investigated experimentally 
using 
optical spectroscopy techniques.
\cite{link1999}
In such measurements, different decay channels can be quantified by the systematic variation of parameters.
\cite{foerster2017}
The dynamics of the particularly intriguing hot electrons can be addressed, at least indirectly,
\latin{via} 
time-resolved spectroscopy techniques.
\cite{Harutyunyan2015, Mjard2016, Heilpern_2018_NatureCommunications}

A more direct access 
to the excited electron ("hot carrier") dynamics 
on the femtosecond timescale 
and in the single-particle limit 
can be gained by the 
time-resolved 2-photon-photoemission (tr-2PPE) technique.
\cite{Fujimoto1984, Schoenlein1988, Schmuttenmaer1994, petek1997, Bauer2015319}
This method 
has been established 
since many years 
for the case of optical rather than plasmonic excitation, 
even in the context of hot carrier assisted photochemistry.
\cite{ma1996}
On the other hand,
plasmonic fields themselves 
have been successfully 
imaged 
with 
time-resolved Photoemission Electron Microscopy (tr-PEEM),
\cite{Schmidt_2002_ApplPhysB} 
both for localized \cite{Cinchetti2005peem, bayer2008time, Lemke_2014_NanoLetters} 
and propagating \cite{Lemke_2014_NanoLetters, Kahl2014, Kahl_2017_Plasmonics} 
surface plasmons (LSP and SPP, respectively).
In this context, 
the photoelectrons were used 
merely 
as an experimental observable for 
the plasmon field. 
Combining these two approaches 
of photoemission
opens up ways 
towards imaging 
plasmon-induced hot carriers 
on the femtosecond and nanometer scale.
However, 
the progress in this direction has been slow, 
as the question 
about
the microscopic mechanism of 
surface plasmon damping 
arose 
two decades ago 
from
early photoemission experiments
on plasmonic samples.
\cite{Scharte2001} 
The separation 
of plasmon and electron dynamics
was limited to 
describing the plasmon
as a modified electromagnetic field 
in the vicinity
of the surface,
\cite{Merschdorf_2004_Phys.Rev.B}
known as the plasmonic near-field.

In recent surface plasmon-induced photoemission experiments,
\cite{Podbiel2017, Lehr_2017_NanoLetters, Tan_2017_NaturePhotonics, Lehr_2019_TheJournalofPhysicalChemistryC, Pettine_2021_ACSnano, Shibuta_2021_ACSNano} 
it is still not easy to distinguish
the plasmon effect from that of the driving laser,
much less the field enhancement effect 
from the more fundamental plasmon decay.
To acquire a photoemission signal dominated
by a plasmonic field,
a strong field-enhancement was used, 
either by using LSP
resonances at nanostructures, 
\cite{Lehr_2017_NanoLetters, Tan_2017_NaturePhotonics, Lehr_2019_TheJournalofPhysicalChemistryC, Pettine_2021_ACSnano, Shibuta_2021_ACSNano}
or 
\latin{via} 
the strong focusing
of an SPP wave.
\cite{Podbiel2017}
Photoemission with contribution 
of an unfocused SPP wave
is present in tr-PEEM,
but its imaging 
requires the interference
with a probe laser pulse.
\cite{Kahl_2017_Plasmonics}
When the plasmon and the probe laser field coexist
in space and time, 
a manifold of possible transition pathways 
with arbitrary, indistinguishable contributions of the two fields 
take part in 
photoexcitation into a state of given energy.
\cite{Spektor_2019_PhysicalReviewX}
This
obscures the experimental access 
to the plasmon-induced hot electrons 
sought for.

In this paper, 
we employ 
two-color time- and energy-resolved PEEM 
experiments 
in real and momentum space 
to directly image plasmon-induced hot carriers. 
We demonstrate the separation of the photoelectrons 
generated in the process, 
in which first a plasmon decays
by producing an excited (hot) electron, 
which is then photoemitted by a photon of double the energy.
To achieve this, 
we use the spatiotemporal dynamics 
of a propagating SPP plane wave pulse.
We perform a theoretical analysis of 
the flow of SPP energy density
to separate
the different contributing terms
of SPP and laser
analytically.
Our method  
provides a direct imaging of plasmon-induced hot electrons,
with sensitivity to the time, space, momentum, and energy domains.

\section{Results and Discussion}
\subsection{Spatiotemporal Separation Scheme}

\begin{figure*}
	\includegraphics[width=\linewidth]{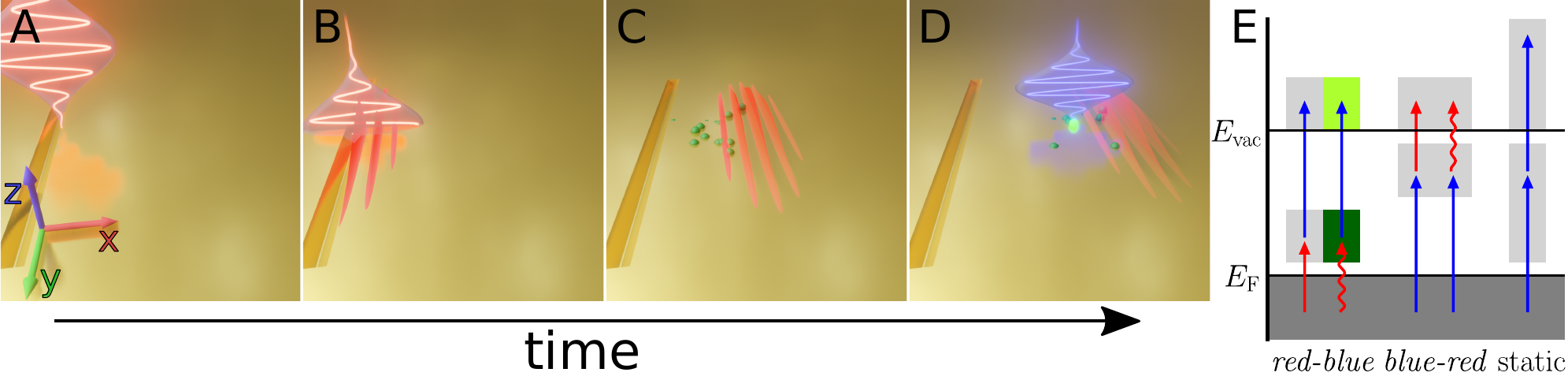}
	\caption{
		Scheme of spatiotemporal separation. 
		A) The pump pulse (red) arrives at the coupling slit etched into the Au surface.
		B) An SPP pulse is excited and interacts with the remaining pump field during the pulse duration.
		C) The plasmon pulse propagates along the surface, 
		exciting hot electrons (green balls) as it is damped through internal decay.
		D) The plasmon-excited hot electrons are photoemitted by a probe pulse (blue). 
		A photoelectron (highlighted green ball) is produced, 
		which is subsequently extracted and detected by the PEEM optics.
		The frames were 
		taken from an animated video of the sequence, 
		provided in the
		supplementary information.
		E) Energy diagram of possible 2PPE channels.
		Transitions are marked as arrows
		representing the driving force,  
		straight red for the pump pulse,
		curly red for the SPP,
		and straight blue for the probe pulse.
		The relevant energy ranges
		of excited 
		and photoemitted
		states 
		are marked 
		in green
		for the 
		plasmon-induced hot carriers
		under investigation,
		or in light gray,
		respectively,
		for the other channels.
		The dark gray area
		represents occupied electronic states 
		up to $E_\mathrm{F}$.
		The green highlighted channel 
		is 
		isolated 
		by means of 
		the spatiotemporal separation scheme.
		}
	\label{fig:propagation_scheme}
\end{figure*}

Our experimental scheme 
to separate the photon- and plasmon-induced hot carriers
is sketched in Figure \ref{fig:propagation_scheme}. 
An ultrashort SPP pulse is excited at a coupling slit,
engraved into a gold layer, 
oriented along the $y$-direction
at \mbox{$x = \SI{0}{\micro\meter}$}.
The red pump pulse 
($\tau \approx \SI{23}{\femto\second}$, $\lambda_{center} = \SI{800}{\nano\meter} \Rightarrow h\nu = \SI{1.55}{\electronvolt}$)
is focused on the coupling structure
under near-normal incidence,
with 
linear polarization in $x$-direction,
perpendicular to the slit.
The structure 
constitutes a break of translation symmetry, 
providing 
additional momentum
which allows the coupling 
of light into the SPP mode.
In that way, 
an SPP pulse with a plane wavefront is launched, 
which then propagates along the Au-vacuum interface 
in positive $x$-direction.
\bibnote{
A corresponding pulse 
propagating in negative $x$-direction
is also launched from the slit,
but it is outside of the detection area.}

The damping  
of the SPP
is dominated by the internal decay in the metal,
producing electron-hole pairs,
because of the nature of the SPP being a dark mode 
that cannot decay into the far-field without a break of symmetry at the surface.
In this way, 
the SPP pulse acts as 
a propagating plasmonic source 
of hot electrons.

We probe the hot electron population 
by a time-delayed irradiation 
of the sample surface 
with 
a blue 
($\lambda_{center} = \SI{400}{\nano\meter} \Rightarrow h\nu = \SI{3.1}{\electronvolt}$)
laser pulse, 
generated by second harmonic generation 
from a split-off part of 
the fundamental output 
of the laser light source.
Using a sub-monolayer of Cesium 
evaporated onto the sample, 
the work function of the Au surface was reduced
to $\Phi \approx \SI{3.4}{\electronvolt}$.
In this way, 
the hot electrons with
an excitation energy of  
$E - E_\mathrm{F} \geq \SI{0.3}{\electronvolt}$
can be photoemitted by absorbing a photon from the probe pulse.
A Photoemission Electron Microscope (PEEM)
is used to image these electrons 
from the sample 
to a 
detector
which is sensitive to their
position and energy
(see Methods).
In this way, 
a 4-dimensional dataset 
of photoelectron yield 
$Y(x,y,E,\Delta t)$
was recorded
by scanning the delay times between the pump and probe pulses 
and acquiring x-y- and energy-resolved PEEM images 
for each time step. 

The possible 
excitation channels
which contribute 
to the measured photoelectron yield
are shown in 
Figure \ref{fig:propagation_scheme}E.
In the dataset, 
the relative yield of the static 2PPE 
by the probe pulse 
("static" channel)
can be referenced and subtracted,
and the dynamic signal
with contribution of excitations with \SI{1.55}{\electronvolt},
namely from the pump or SPP pulses, 
remains.
For this signal, 
there are two participating channels,
which differ in 
the temporal order of 
the two sequential excitation steps.
Apart from the 
previously outlined 
\textit{red-blue} channel,
in which the excited electron 
from the red pump pulse or SPP
is photoemitted by the blue probe,
the reversed order
is also possible,
in which the blue probe laser 
excites an electron 
which is then photoemitted
by the red pump laser or SPP
(\textit{blue-red} channel).
The intermediate state of 
the \textit{blue-red} channel
is located at an energy
\SI{1.55}{\electronvolt}
higher than the one of the \textit{red-blue} channel though.
This state has an
about ten times
shorter inelastic lifetime,
\cite{Bauer2015319}
following essentially  
the well-known relation
$T_1 \propto (E-E_\mathrm{F})^{-2}$
from Fermi liquid theory.
Neglecting band structure effects, 
the 2PPE yield 
scales at least linearly 
\bibnote{
The exact scaling depends 
on dephasing of the coherences 
along the excitation pathway.}
with 
$T_1$,
therefore, 
the \textit{red-blue} channel
dominates the signal.

The further isolation 
of the contribution 
of the plasmon-induced hot carriers
from the ones excited directly by the pump laser 
is performed using 
the spatiotemporal signature
of their source,
namely
the SPP pulse, 
which propagates along the surface of the sample on a femtosecond time scale.
To characterize the spatiotemporal characteristics of this source
as a reference for the following evaluation of the photoemission data, 
we analytically calculated 
the energy density flow of the time-dependent electromagnetic field 
in a model system that emulates the conditions in the experiment,
but without a probe laser pulse
(see Methods).

We have analytically derived 
the absorbed energy density rate 
by solving Maxwell's equations 
for the laser and SPP fields. 
This calculation describes 
the laser energy absorption 
and defines the spatiotemporal characteristics 
of the available energy 
for the excitation of hot electrons.
We have shown in
Ref.  
\citenum{Terekhin_2020_AppliedSurfaceScience} 
that it contains different interference contributions 
(\laserlaser, \laserspp, and \sppspp). 
Each of these terms
contributes 
to the first step 
of the 
\textit{red-blue}
photoemission channel
in Figure \ref{fig:propagation_scheme}E,
producing the hot electrons
in the intermediate state of the process.
The first term,
\laserlaser,
has no plasmonic contribution and 
is responsible for
purely photon-induced electrons. 
The second term,
\laserspp,
arises from the interference of the laser and SPP fields.
We have shown in 
Ref.
\citenum{Spektor_2019_PhysicalReviewX}
with a one-color phase-resolved PEEM experiment
that this term produces mixed contributions
in the photoemission process 
where even entangled quantum pathways take part,
in which both fields participate in an indistinguishable manner.
This corresponds 
to a mixing between 
the two 
\textit{red-blue} channels 
displayed 
in Figure \ref{fig:propagation_scheme}E.
Therefore, 
a hot electron produced by the \laserspp~term 
cannot be clearly classified as 
photon-induced or plasmon-induced. 
However,
the third term, 
\sppspp, 
describes the propagation and decay 
of the SPP pulse. 
This term is the origin of 
purely plasmon-induced hot carriers. 
A detailed description of 
the calculation and
the resulting energy density rate
in space and time  
is given in our previous work.
\cite{Terekhin_2020_AppliedSurfaceScience}

\begin{figure*}
	\includegraphics[width=\linewidth]{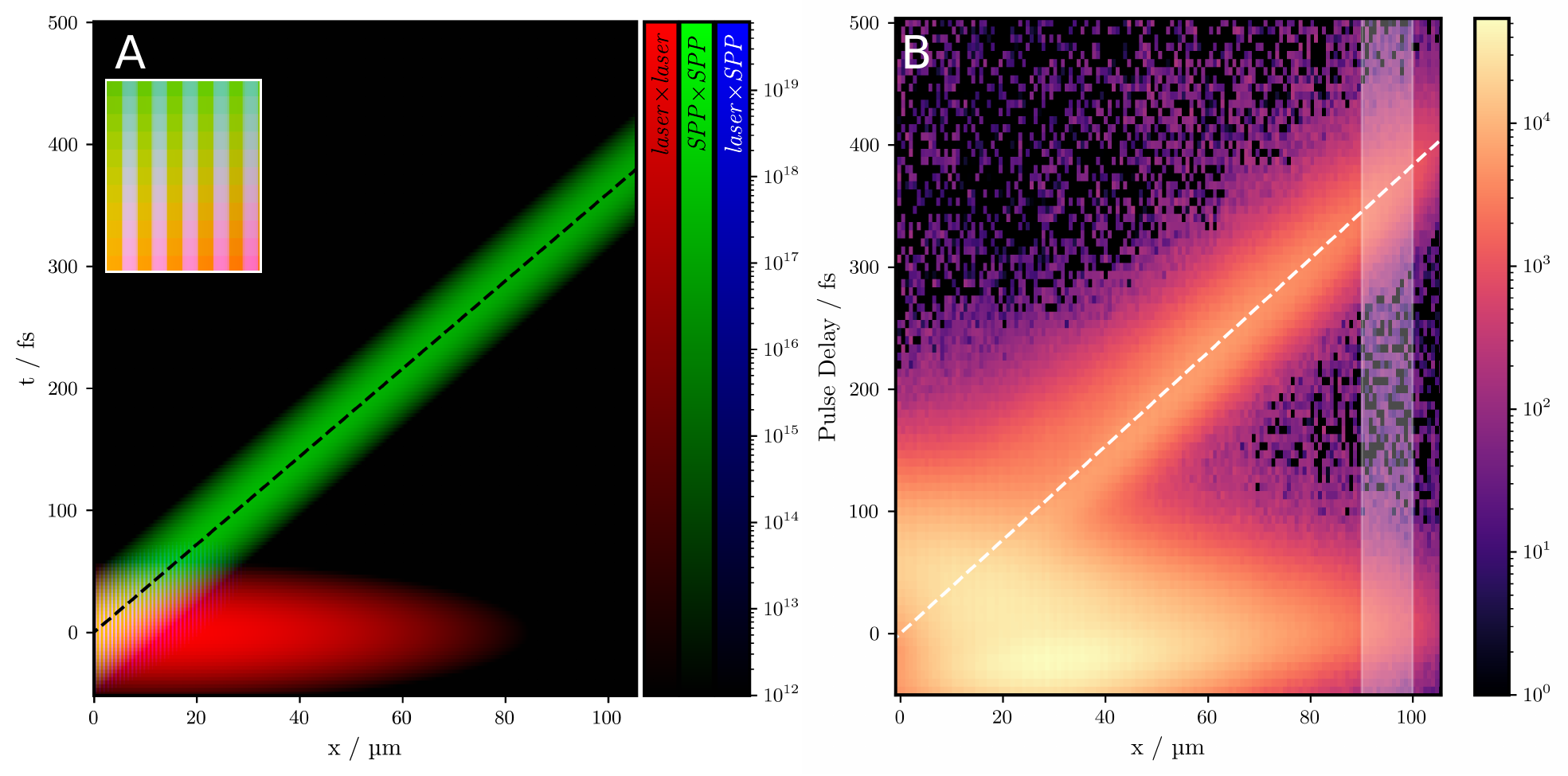}
	\caption{
		A) 
		Time-dependent energy density rate
		of the electromagnetic field contributions 
		at the surface,
		for a coupling structure
		at \mbox{$x = \SI{0}{\micro\meter}$} 
		irradiated at normal incidence
		with a
		pump pulse 
		centered at $t = \SI{0}{\femto\second}$.
		The three contributing terms, 
		\laserlaser, \sppspp, and \laserspp~are 
		encoded in the 
		red, green, and blue
		color channels of the image,
		with each colorbar given in 
		\si{\J \per \second \per \cubic \meter}.
		The additive color mixing
		leads to additional color shades 
		where more than one field is present
		at the same time and position,
		which leads to
		yellow pixels for \laserlaser~and \sppspp,
		and more blue-shaded colors for all three components.
		The velocity of an SPP $v_\mathrm{SPP}$ excited with $\lambda = \SI{800}{\nano\meter}$ is plotted as a dashed black line.
		Plots of the single components are available in the supplementary information.
		The inset is a zoom of 
		$x =$ \SIrange{10}{15}{\micro\meter}
		and
		$t =$ \SIrange{15}{45}{\femto\second},
		where all three components are present
		and the contribution of the interference term 
		\laserspp~is 
		visible as blue-shaded stripes.
		B) 
		Measured tr-PEEM signal 
		of the
		spatiotemporal signature of hot electrons,
		$Y_{\text{source}}$,
		with the colorbar given in counts,
		at an excitation energy of  
		\SIrange{1.45}{1.55}{\electronvolt},
		for an integration range of $y = $\SIrange{55}{79}{\micro\meter}. 
		The static background $Y_{\text{static}}$ was subtracted.
		The perceived velocity of the SPP pulse $v_{\mathrm{perceived}}$ 
		(see Methods)
		is plotted as a white dashed line.
		The white shaded region represents the $x$ integration range for 
		Figure \ref{fig:delaytoenergy}.
	}
	\label{fig:xtdata}
\end{figure*}

The spatiotemporal dependence of the energy density rate 
at the interface between Au and vacuum 
is shown in Figure \ref{fig:xtdata}A,
with the three 
separate
contributing terms
encoded in the 
red, 
green, 
and blue
color channels,
representing 
\laserlaser, 
\sppspp, 
and \laserspp, 
respectively. 
The pump pulse 
(red channel, \laserlaser~term)
is visible as a non-propagating signal around 
$t=\SI{0}{\femto\second}$.
The SPP 
(green channel, \sppspp~term)
appears
as a propagating trace 
in positive $x$-direction 
until well past the irradiation time. 
The interference term
(blue channel, \laserspp~term)
is present only 
where the two fields
overlap in space and time,
leading to a periodic variation
from yellow to blue-shaded colors
(inset in Figure \ref{fig:xtdata}A).

The dashed black line in Figure \ref{fig:xtdata}A represents the SPP group velocity. 
This velocity can be derived from the SPP dispersion relation as 
\begin{equation}
	v_\mathrm{SPP} = \frac{d \omega}{d k^\prime_\mathrm{SPP}}
	\text{,}
\end{equation}
where $k^{\prime}_\mathrm{SPP}$ is the real part of the SPP wave vector
\begin{equation}
	k_\mathrm{SPP} 
	= \frac{\omega}{c_0} \cdot \sqrt{\frac{\epsilon_\mathrm{m} \epsilon_\mathrm{d}}{\epsilon_\mathrm{m} + \epsilon_\mathrm{d}}}
	\text{,}
\end{equation}
where 
$\omega$ is the laser angular frequency,
$c_0$ is the speed of light, 
and $\epsilon_\mathrm{m}$ and $\epsilon_\mathrm{d}$ are the dielectric functions of the metal and dielectric half-spaces, respectively. 

\subsection{PEEM Results\label{sec:isolation}}
In the 
time-dependent energy density rate 
in Figure \ref{fig:xtdata}A,
the laser and SPP contributions are clearly separated 
in time 
at larger distances from the excitation edge 
($x \gtrsim \SI{40}{\micro\meter}$).
During 
the pulse duration $\tau$,
when 
the pump pulse impinges on the sample 
around 
$t = \SI{0}{\femto\second}$,
the \laserlaser~term
dominates.
An electron which is photoemitted at this stage 
close to the excitation edge 
cannot be attributed unambiguously
to either photonic or plasmonic excitation.
In contrast,
after the pulse has faded for 
$t > \tau$, 
the situation is different:
The SPP pulse still propagates along the surface 
as a moving source, 
producing hot electrons
along its way.
An electron 
which is 
photoemitted 
and detected
during this stage 
might still remain from a photonic excitation,
but especially for higher energies 
where electron lifetimes are as short as only a few tens of \si{\femto\second},
the SPP pulse 
is the predominant source of hot electrons.

The experimental PEEM data
can be analyzed 
in terms of 
the same space-time characteristics
to assign types of origin 
to the detected electrons.
From the
acquired 4D dataset
of photoemission yield
$Y(x,y,E,\Delta t)$,
the 
contribution 
caused by the calculated dynamic source,
$Y_{\text{source}}$,
is isolated
by subtracting the
static (delay-independent) 2PPE yield:
\begin{align}
&Y_{\text{source}}(x,y,E,\Delta t) \nonumber\\ 
&= Y(x,y,E,\Delta t) \nonumber\\
&- {Y_{\text{static}}(x,y,E)}
\text{.}
\label{eq:bluenorm}
\end{align}
The static yield 
$Y_{\text{static}}$ 
is derived from the relative signal  
before the two pulses overlap,
by averaging 
the measured count values 
for the pulse delays
in the range of 
$\Delta t < \SI{-200}{\femto\second}$.
The only 
delay-independent
2PPE channel
(\textit{static} in Figure \ref{fig:propagation_scheme}E)
is
the two-photon contribution 
from the probe pulse 
(\textit{blue-blue}).
It is the dominant contribution to
$Y_{\text{static}}$.
Additionally,
a three-photon contribution from the pump pulse 
("Red 3PPE")
is present near the coupling slit, 
where the field of the pump pulse is strong,
but 
its photoemission yield is small
due to the higher-order photoemission process
needed 
to overcome the work function.
In the range of spatiotemporal separation ($x \gtrsim \SI{40}{\micro\meter}$), 
it is negligible.
Real-space plots of $Y_{\text{static}}$ are available in the supplementary information.

The resulting dynamic photoemission yield 
$Y_{\text{source}}(x,y,E,\Delta t)$
is shown in Figure \ref{fig:xtdata}B
as a 2D projection to the 
$x$ - $\Delta t$ - plane.
The 
integration range of
$y =$ \SIrange{55}{79}{\micro\meter}
corresponds to 
the irradiated section of the coupling slit.
The 
integration range 
in energy
of 
$E - E_\mathrm{F} = $\SIrange{1.45}{1.55}{\electronvolt}
corresponds to electrons 
excited from close to the Fermi edge.
For these
highest available excitation energies,
the inelastic lifetime of hot electrons is shortest,
in the order of tens of femtoseconds,
\cite{Bauer2015319}
and the trace is the least broadened in the time domain.
Therefore, 
this range is best suited
for comparison 
with 
Figure \ref{fig:xtdata}A 
to assign regions of 
predominant photonic and plasmonic excitation
for the analysis 
of the 
respective
electron energy distributions.

The obtained hot electron trace 
in Figure \ref{fig:xtdata}B
can be clearly attributed 
to the spatiotemporal characteristics of the source: 
After an intense non-propagating feature 
that is
caused by the pump pulse,
a trace of electrons excited by the SPP is observed.
For larger delays and distances from the coupling slit, 
the photon- and plasmon-induced hot electrons are clearly separated.
The electrons observed 
in the propagating trace 
after the pump pulse has faded
can be assigned to the 
photoemission channel
in which the SPP contributes the initial excitation
(green highlighted path in Figure \ref{fig:propagation_scheme}E).
This signal 
serves as an observable for 
the plasmon-induced hot electrons
which are the main focus of this study.

Comparing the experimental PEEM trace 
to the 
temporal profile of the calculated energy density rate
(Figure \ref{fig:xtdata}B vs. \ref{fig:xtdata}A),
the influence 
of the inelastic electron lifetime 
is visible in the slower temporal attenuation 
of the PEEM signal.
An additional effect of broadening 
is added by 
the finite temporal length of the probe pulse,
which universally provides a lower limit to 
the temporal width of a pump-probe signal,
given by the cross-correlation between the two pulses.

\begin{figure}
	\includegraphics[width=\linewidth]{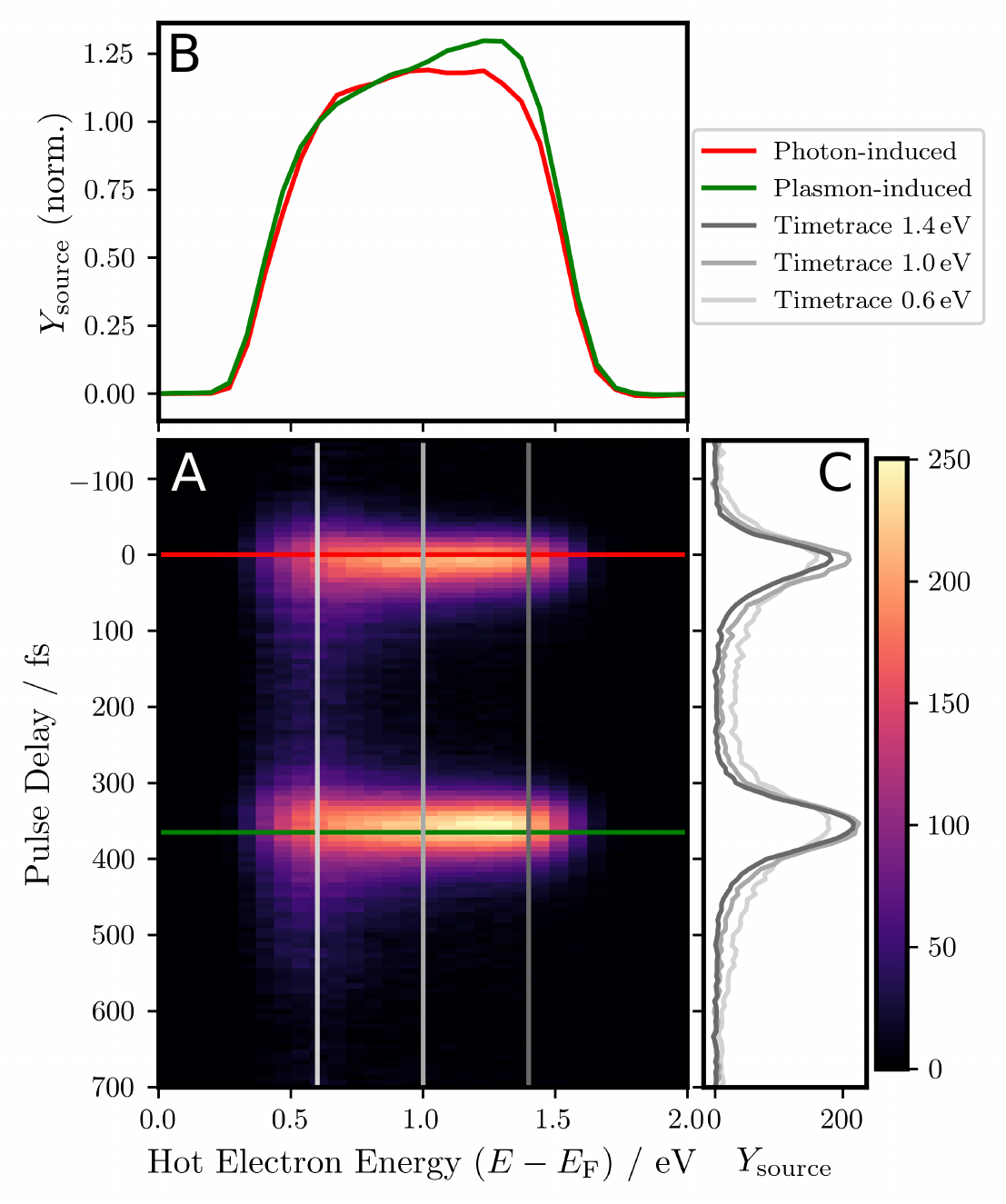}
	\caption{
	A)
	Projection of the 4D-dataset 
	of photoemission yield 
	for 
	$x =$ \SIrange{90}{100}{\micro\meter}, 
	$y =$ \SIrange{55}{79}{\micro\meter}
	to the time-energy plane.
	B)
	Spectra 
	of photon-induced (red line)
	and plasmon-induced (green line)
	hot electrons,
	extracted along the respective lines
	in A)
	at
	$\Delta t_0 = \SI{0}{\femto\second}$ (red) 
	and $\Delta t_\text{SPP} = \SI{365}{\femto\second}$ (green),
	and normalized 
	to 
	the onset of the
	low-energy cutoff 
	at
	$E-E_\mathrm{F} = \SI{0.6}{\electronvolt}$.
	C)
	Time trace of electrons 
	emitted  
	from excitation energies of
	$E - E_\mathrm{F} = $ 
	\SI{1.4}{\electronvolt},
	\SI{1.0}{\electronvolt},
	and \SI{0.6}{\electronvolt},
	extracted as cuts along the respective gray lines
	in A).
	The colorbar and $Y_\text{source}$ axis values are given in counts.
	}
	\label{fig:delaytoenergy}
\end{figure}

For the study of 
plasmon-induced hot electrons,
and in contrast to
previously reported two-color PEEM experiments with SPP,
\cite{Joly_2018_TheJournalofPhysicalChemistryC}
the key in our isolation scheme is the combination of information 
of all available dimensions,
in our case time, space, and energy.
In the full 4-dimensional dataset, 
different aspects of spectral and lifetime information 
can be extracted by evaluating different regions of the hypercube.
This can be seen 
in the projection of certain slices of the dataset:

In the range of $x = \SI{95}{\micro\meter}$, 
where the photon- and plasmon contributions are most clearly separated
in space and time 
(white shaded range in Figure \ref{fig:xtdata}B),
a projection of the experimental 4D dataset 
to the $E$-$\Delta t$-plane
was plotted in Figure \ref{fig:delaytoenergy}A.
Line cuts were extracted
using a temporal integration range of $\pm \SI{25}{\femto\second}$
and an energy integration range of $\pm \SI{0.05}{\electronvolt}$.
These cuts represent 
electron spectra for specific times (green, red, Figure \ref{fig:delaytoenergy}B)
and time traces for constant energies (gray, Figure \ref{fig:delaytoenergy}C),
respectively.

In the time traces
(in Figure \ref{fig:delaytoenergy}C), 
the typical lifetime behavior of hot electrons,
governed by the Fermi liquid theory,
\cite{Bauer2015319}
is visible:
The inelastic lifetimes 
$T_1$
increase with smaller excitation energy,
which leads to more electrons remaining for later times after excitation
(lighter gray compared to darker gray graphs).

In the energy domain,
a photon-induced and a plasmon-induced hot electron spectrum 
is plotted for delay times of
$\Delta t_0 = \SI{0}{\femto\second}$ (red) 
and $\Delta t_\text{SPP} = \SI{365}{\femto\second}$ (green),
representing the stages at which 
the respective photonic and plasmonic contributions
to the source field are most dominant.
The spectra
in Figure \ref{fig:delaytoenergy}B
are normalized 
to their low-energy cutoff
(for absolute values see raw spectra in the
supplementary information).
A comparison shows 
that the higher excitation energies 
appear stronger in the plasmon-induced spectrum
(see also 
difference spectrum
in the
supplementary information).
This apparent preference 
for higher-energetic excitation
is intriguing:
Given that 
the photonic signal 
follows the familiar excitation mechanism
with the energy distribution 
governed by the DOS of the material,
it suggests that for the plasmonic signal
a different excitation mechanism
is involved.
The high-energy feature
emerging among
the still present, 
continuous spectrum of 
photon-induced electrons
hints
that both the familiar single-particle excitation
and a different plasmonic excitation
might contribute.
This would reflect the hybrid nature
of the surface plasmon polariton
as 
an electromagnetic wave (\textit{polariton} aspect)
bound to
a collective charge motion (\textit{plasmon} aspect).
Following this reasoning,
the \textit{polariton} aspect 
would cause an Einsteinian photoexcitation known from light,
while the \textit{plasmon} aspect
selectively excites electrons
close to $E_\mathrm{F}$,
similar to the effect known from bulk plasmons.
\cite{Hopfield_1965_PhysicalReview, Barman_2004_SurfaceScience, Barman_2004_PhysicalReviewB, Reutzel_2019_PhysicalReviewLetters, Li_2021_ACSPhotonics, Novko_2021_PhysicalReviewB}

An influence of 
the weaker \textit{blue-red} channel
should manifest in the time-resolved signal 
predominantly at time steps 
where the blue pulse precedes the red,
for 
$\Delta t \lesssim \Delta t_0$
or 
$\Delta t \lesssim \Delta t_\text{SPP}$,
owing to the order of the involved excitation processes.
Especially in the lower energy channels,
secondary electrons 
would be expected
due to the shorter lifetime
of the intermediate state
as discussed above.
Such an influence
is not 
observed 
in the 
electron 
traces,
which leads us to conclude
that the channel
is not of significant strength,
confirming that 
the observed hot electron distributions
are predominantly excited 
by the pump and SPP pulses.

\subsection{Momentum Microscopy}\label{sec:kspace}

\begin{figure*}
	\includegraphics[width=\linewidth]{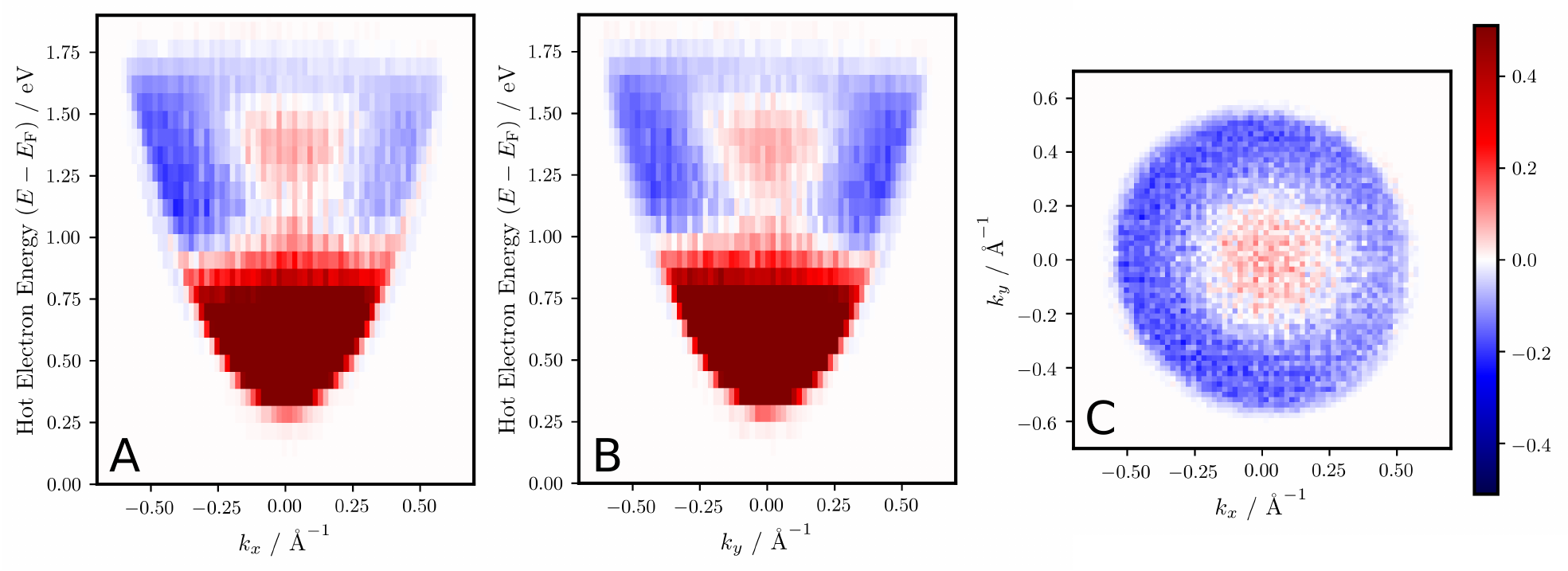}
	\caption{
		Momentum distribution of plasmon-induced hot electrons relative to photon-induced hot electrons.
		Cuts
		of the normalized difference values $\Delta Y(k_x, k_y, E)$
		are shown 
		along 
		A) the $E$-$k_x$-direction for $k_y = \SI{0}{\per \angstrom}$, 
		B) the $E$-$k_y$-direction for $k_x = \SI{0}{\per \angstrom}$, 
		and C) the $k_x$-$k_y$-direction for $E - E_\mathrm{F} = $\SIrange{1.4}{1.5}{\electronvolt}.
		}
	\label{fig:kspacedata}
\end{figure*}

To gain further insight 
into the intriguing feature
of high-energy preference
of SPP-induced hot electrons,
the momentum microscopy mode of the PEEM 
\cite{Kotsugi_2003_ReviewofScientificInstruments, Tusche_2016_AppliedPhysicsLetters, Haag_2019_ReviewofScientificInstruments, Maklar_2020_ReviewofScientificInstruments}
was used to characterize
the distribution of 
plasmon-induced hot electrons 
in 
energy and 
momentum space ($k$-space).

To use the concept of spatiotemporal separation 
in this mode,
the probe laser focus was 
positioned in 
the center of the field-of-view, 
\SI{\sim 100}{\micro\meter} to the right of
the coupling slit,
and an iris aperture 
in the image plane of the electron-optical column
was closed to limit the detection area
to a diameter of \SI{\sim 30}{\micro\meter} in real space.
In this way,
the SPP pulse 
propagates through the selected area
after the pump pulse is no longer present on the sample,
similar to the temporal evolution
in the white-shaded area of
\mbox{Figure \ref{fig:xtdata} B}.

The momentum space distribution of the emitted electrons
was detected 
by imaging the back-focal plane 
of the objective lens of the PEEM 
to the detector,
thus acquiring a $k$-space 
photoemission dataset 
of the shape
$Y(k_x,k_y,E,\Delta t)$.
Similarly to 
the subtraction of the static yield  
in
equation (\ref{eq:bluenorm}),
the signal contribution 
caused by the SPP source field 
(green highlighted path in Figure \ref{fig:propagation_scheme}E)
was extracted
using the delay time when the SPP pulse is centered on the selected area,
$\Delta t_\text{SPP} = \SI{360}{\femto\second}$,
and
subtracting the static yield before the temporal overlap at
$\Delta t_\text{static} = \SI{-334}{\femto\second}$:
\begin{align}
&Y_{\text{SPP}}(k_x,k_y,E) \nonumber\\
&= Y(k_x,k_y,E,\Delta t_\text{SPP}) \nonumber\\
&- Y(k_x,k_y,E,\Delta t_\text{static})
\text{.}
\label{eq:bluenorm_kspace}
\end{align}

A reference momentum distribution 
for photon-induced electrons 
$Y_\text{photon}$
(grey in \textit{red-blue} channel in Figure \ref{fig:propagation_scheme}E)
was measured
with the red pump laser 
also centered to the field-of-view instead of focused on the excitation slit,
using identical sample position and PEEM settings.
The pump-induced signal contribution was extracted
at time zero 
$\Delta t_0 = \SI{0}{\femto\second}$,
subtracting the static probe yield  
in the same way:
\begin{align}
&Y_{\text{photon}}(k_x,k_y,E) \nonumber\\
&= Y_\text{ref}(k_x,k_y,E,\Delta t_0) \nonumber\\
&- Y_\text{ref}(k_x,k_y,E,\Delta t_\text{static})
\text{,}
\label{eq:bluenorm_kspace_photon}
\end{align}
where $Y_\text{ref}$ is the photoelectron yield of the reference measurement.

Figure \ref{fig:kspacedata} shows the normalized difference $\Delta Y(k_x, k_y, E)$
between plasmon-induced and photon-induced hot electrons 
in momentum space.
The values 
were normalized 
by 
\begin{equation}
\Delta Y(k_x, k_y, E) = 
\frac{Y_\text{SPP}}{\langle{Y_\text{SPP}}\rangle} 
- \frac{Y_\text{photon}}{\langle{Y_\text{photon}}\rangle}
\textbf{,}
\label{eq:normdiff}
\end{equation}
where 
$\langle{Y_\text{SPP}}\rangle$ and $\langle{Y_\text{photon}}\rangle$
are the respective mean values 
of $Y_\text{SPP}$ and $Y_\text{photon}$ 
over the full energy and momentum range,
to account for the arbitrary difference 
in absolute signal strength.
Plots 
of the same $k$-space cuts 
of the individual yields  
$Y_\text{SPP}$ and $Y_\text{photon}$ 
are provided in the
supplementary information.

The high-energy feature 
described in the previous section
is present 
in the $k$-space data 
in Figure \ref{fig:kspacedata}
as a contiguous red region
in the energy range of 
\SIrange{1.2}{1.55}{\electronvolt}
in the center of the momentum range,
close to the $\overline{\Gamma}$-point of the surface.
In this region, 
the value of 
$\Delta Y(k_x, k_y, E)$
is positive, 
corresponding to an enhanced generation of hot electrons
by plasmons with respect to photons.

From the usual perspective of
surface science,
where one typically explains features in $k$-space
in terms of the electronic band structure,
this feature is surprising:
Due to the polycrystalline nature
of the sample
and the consequent randomness of crystallite orientations
contained in the detection area, 
band structure effects 
should average out
and
one expects 
a homogeneous momentum distribution of electrons. 

Interestingly,
no significant 
anisotropy
or
asymmetry 
of the plasmon-induced electrons
along the propagation direction of the SPP
($k_x$-direction) is observed.
The enhancement is 
concentrated at 
small values of the in-plane momenta 
$k_\parallel = (k_x^2 + k_y^2)^{1/2}$,
but otherwise
symmetric in $k$-space.
This implies that
the wave vector orientation and related momentum
of the SPP wave 
seems to have
no relevant influence.

The obtained momentum distribution 
could be an important clue 
for a different mechanism
taking part in
electron excitation by surface plasmons.
The transfer of energy from
the collective to single-particle excitation
seems to efficiently couple 
to electrons with 
low momentum
and close to the Fermi surface.
Possible origins of this phenomenon
are to be looked for in the nature of the SPP:
Apart from the difference in the orientation 
of oscillating  electric and magnetic fields,
the movement of electrons as part of the collective excitation
(\textit{plasmon} aspect) 
in contrast to the single-particle case
could be crucial.
But ultimately,
at this point 
a conclusive explanation
of the plasmon decay mechanism, 
which links the collective motion 
to the single particle momentum,
is missing.

In Figure \ref{fig:kspacedata} A and B,
which show the normalized difference  
in the energy vs. momentum distribution
in both directions 
($E$-$k_x$-plot and $E$-$k_y$-plot),
a strong enhancement for 
\textit{smaller} energies
of 
$E - E_\mathrm{F} < \SI{0.8}{\electronvolt}$
is also apparent,
which seems to contradict the 
result of the real space experiment.
It turns out though,
that this is an artifact
of the measurement parameters:
In contrast to the pump pulse,
which impinges on the sample under near-normal incidence
and is present only for the laser pulse duration,
the SPP pulse propagates with finite velocity
through the detection area 
selected by the iris aperture.
This results in an apparent broadening
of the SPP pulse duration,
increasing the detection probability
for secondary electrons,
which are produced by inelastic decay 
after the initial excitation.
We show in the 
supplementary information
that this effect can be mitigated
using a smaller aperture size
and extracting the data from 
time windows of the full time-resolved measurement,
selected to compensate 
for the apparent broadening.
Unfortunately, 
the use of a very small aperture causes a different artifact
in the electron-optics,
therefore,
we chose to show the present data in the main manuscript.

\section{Conclusion}
To summarize,
we have shown how 
direct experimental access to plasmon-induced hot electrons,
separated from the influence of 
driving laser and field enhancement effects,
was realized 
in a photoemission experiment.
This provides a novel opportunity for 
the exploration of plasmon-enhanced 
energy conversion and chemistry
from the fundamental perspective of solid-state physics.

The results 
show a preference 
for higher-energetic excitations 
with small in-plane momentum
induced by surface plasmons in contrast to photons,
independent of the crystal orientation 
of the metal surface.
This suggests that the mechanism 
of plasmon-enhanced energy harvesting
is more fundamentally linked 
to the nature of the surface plasmon
and its decay into hot carriers,
rather than just a simple field-enhancement
at the metallic surface.
At this point 
it is unclear
if the observed effect
is of related physical origin 
as the similar-looking effect 
known for bulk plasmons.
\cite{Hopfield_1965_PhysicalReview, Barman_2004_SurfaceScience, Barman_2004_PhysicalReviewB, Reutzel_2019_PhysicalReviewLetters, Li_2021_ACSPhotonics, Novko_2021_PhysicalReviewB}
The aspect of a
collective oscillation of electrons
in both bulk and surface plasmons
leads to the hypothesis 
that a similar mechanism
may be at play.
On the other hand,
it must be noted 
that many aspects are fundamentally different:
While the bulk plasmon 
is in its nature
a resonance effect of the bulk electronic system,
the SPP case 
constitutes a 
decay of surface plasmons 
at the metal-dielectric interface
and 
far from resonance.
In both cases,
the preferential excitation 
of electrons at $E_\mathrm{F}$
is evident,
but as the SPP 
follows the frequency of the pump light
rather than being pinned 
to a resonance frequency,
the non-Einsteinian characteristic
of subsequent pinning of electron energy
off from $\hbar \omega$ of the driving laser
is \emph{not} to be expected
for the SPP case.

Irrespective of fundamental origin,
a preferential generation 
of high-energy electrons
holds great potential
for chemical and energy harvesting purposes.
Unlike the 
typical continuous distribution
of energy to electrons and holes,
here we have a case 
in which a concentration
of energy to "hot" electrons
with only low-energetic ("cold") holes
is evident,
which could be of key advantage.
For technical applications,
a promising decay process 
is the chemical interface damping 
\cite{Wu_2015_Science, Kale_2015_Science, Khurgin_2021_ACSPhotonics}
at a metal interface, 
where carriers are directly excited in the adjacent material. 
It is yet to be determined 
whether the observed effect 
could play a role in this context,
but an enhancement of plasmonic excitation 
for electrons at $E_\mathrm{F}$
was also reported for 
plasmonic nanoparticle interfaces.
\cite{Tan_2017_NaturePhotonics, Tan_2018_PhysicalReviewLetters, Shibuta_2021_ACSNano}
On the other hand, 
the electronic structure at the chemical interface 
in such systems
additionally provides emerging dephasing pathways, 
which are of importance for plasmon decay,
\cite{Therrien_2019_FaradayDiscussions, Khurgin_2021_ACSPhotonics}
as well as the dynamics of hot electrons at the interface.
\cite{Foerster_2020_NanoLetters}
In fact,
emergent plasmonic excitation features 
were explained in terms of interface effects
in most previous work.
The concept of spatiotemporal separation 
will bring further insight 
to the multitude of influential effects 
for plasmon damping and plasmonic excitation 
at such interfaces 
by making the plasmon-induced carriers directly experimentally accessible.

\appendix
\section{Methods\label{sec:methods}}

\subsection{Time-dependent Energy Density Calculations\label{sec:methods:theory}}
A detailed description of 
the calculation method 
is given in our previous work.
\cite{Terekhin_2020_AppliedSurfaceScience}
In our simulations,
we consider a Gaussian pulse profile for the incoming laser field, 
including all components of electric and magnetic fields. 
The system consists of 
the gold sample filling the half-space $z \leq 0$ 
and a perfect vacuum filling the other half-space $z > 0$.
As a spatially defined coupling structure providing SPP,
we use a step edge perpendicular to the interface of 
height \SI{50}{\nano\meter} 
at \mbox{$x = \SI{0}{\micro\meter}$} 
irradiated at normal incidence. 

The laser parameters applied in the simulations are 
a laser wavelength centered at  
\mbox{$\lambda = \SI{800}{\nano\meter}$}, 
a pulse duration of 
\mbox{$\tau = \SI{23}{\femto\second}$}, 
a Gaussian width of \SI{25}{\micro\meter} 
centered at $x = \SI{10}{\micro\meter}$ away from the position of the step edge 
and a pump pulse energy of \SI{1}{\nano\joule}. 
The pulse arrives at the sample surface 
centered with its maximal amplitude at $t = \SI{0}{\femto\second}$.
The coupling parameter at the step edge, 
$\beta$, 
is set to \SI{0.2}{}. 
It describes the ratio of moduli 
of the SPP magnetic field amplitude
and the incident laser magnetic field amplitude
at the origin of the step edge ($x=0$). 
The obtained energy density rate
was averaged over one full period of the oscillating fields 
to account for the lack of phase resolution
of the experiment.

We represent the dielectric half-space by a perfect vacuum with $\epsilon_\mathrm{d} \equiv \SI{1}{}$.
The metallic half-space is modeled with the dielectric function reported by 
Olmon 
\latin{et al.} 
\cite{Olmon_2012_Phys.Rev.B} 
for evaporated gold.

\subsection{Sample\label{sec:methods:sample}}
The sample was produced in the Nano Structuring Center (NSC) at TU Kaiserslautern. 
A layer of gold
with a thickness of \SI{250}{\nano\meter}
was sputter-deposited onto 
a substrate cut from a native-oxidized Silicon wafer.
The straight coupling slit 
with 
a width of \SI{\sim100}{\nano\meter},
a depth of \SI{\sim230}{\nano\meter},
and a length of \SI{80}{\micro\meter}
was etched by focussed ion beam (FIB) milling.
A sub-monolayer of Cesium 
was evaporated onto the sample surface
\latin{in situ} 
to reduce the work function for the PEEM measurement.

\subsection{Two-color Pump-Probe Setup\label{sec:methods:pumpprobe}}
A femtosecond Ti:Sapphire laser (Spectra Physics Tsunami) produces pulses at a central wavelength of 
$\lambda = \SI{800}{\nano\meter}$
with a pulse duration down to 
$\tau = \SI{23}{\femto\second}$.
In a home-built two-color time-resolved optical setup, 
after using a prism compressor for dispersion compensation,
the pulses are split up in two arms of an interferometer.
In one arm the ("probe") pulse is frequency-doubled with a 
beta barium borate (BBO) crystal
and compressed again with another prism pair,
then it is routed over a linear delay stage to adjust the delay between the pump and probe.
The "pump" arm passes a fixed optical route of equal effective length.
The two arms are subsequently recombined 
and 
irradiated onto the sample in the PEEM setup.

\subsection{Photoemission Electron Microscopy\label{sec:methods:peem}}
We use a customized commercial PEEM setup (IS-PEEM, Focus GmbH),
which is designed for laser irradiation under near-normal incidence.
\cite{Kahl2014}
The laser beam is routed over a small mirror, 
introduced into the electron column of the microscope near the optical axis, 
which leads to an angle of incidence $\text{AOI} \approx \SI{4}{\degree}$.

The beam diameters on the sample are 
approximately 
$\SI{50}{\micro\meter}$,
the pump beam centered at the excitation slit
and the probe beam offset 
to 
$x \approx \SI{70}{\micro\meter}$ for the real space experiments
and 
$x \approx \SI{100}{\micro\meter}$ for the momentum microscopy experiments.

The photoelectrons are detected with a delayline detector (DLD),
which
records their positions as well as their time-of-flight, 
with reference to a trigger signal from the laser.
Thus,
the kinetic energy of each photoelectron is measured
through proper energy-calibration of the detector's time channels,
and a time- and energy-resolved PEEM experiment 
\cite{Oelsner2010317}
is performed.
The PEEM can be operated 
as a momentum microscope
by imaging the back-focal plane 
of the objective lens,
in this way detecting 
the angular distribution of photoelectrons.

In the real-space experiment,
the SPP trace is observed 
through the probe pulse, 
which hits the sample under near-normal, 
but not perfectly normal incidence
of 
$\text{AOI} \approx \SI{4}{\degree}$.
Therefore, 
the in-plane component of the wave vector of the probe pulse
$k_\parallel = - {\omega}/{c_0} \cdot \sin (\text{AOI})$
contributes to the perceived propagation.
For the dashed white line 
in \mbox{Figure \ref{fig:xtdata}B},
this was taken into account by plotting the perceived SPP velocity 
(see derivation in SI)
\begin{equation}
	v_{\mathrm{perceived}} = v_{\mathrm{SPP}} \cdot \frac{1}{1 + \frac{v_\mathrm{SPP}}{c_0} \sin (\mathrm{AOI})}
	\text{,}
\end{equation}
which is slightly slower than the real SPP velocity $v_{\mathrm{SPP}}$.

\subsection{Data Evaluation\label{sec:methods:data}}

For all tr-PEEM datasets, 
several runs of exposures
for the same list of pump-probe delays
were acquired and
and summed up during post-processing 
to improve the data quality.

The energy axis offset of all datasets 
was calibrated to the excited electron energy
by fitting a Fermi distribution 
to the high-energy cutoff of the photoelectron spectrum
acquired with the blue probe laser.
The fitted Fermi energy was set to 
$E - E_\mathrm{F} = \SI{3.1}{\electronvolt}$
according to the photon energy of the laser.

The real space data in 
Figures \ref{fig:xtdata} B and \ref{fig:delaytoenergy}
were binned in groups of
(2, 16) 
pixels in 
($x$, $y$)
and a constant background 
across energy channels 
was subtracted 
for each binned pixel
prior to the subtraction of the static signal in
equation (\ref{eq:bluenorm})
to increase the signal-to-noise ratio.

The $k$-scale in the momentum space measurements
(Figure \ref{fig:kspacedata})
was calibrated by fitting a parabolic free-electron dispersion
to the low-energy cutoff of the data
in $E$-$k_x$- and $E$-$k_y$-cuts of the center of the data.

For the calculation in equation (\ref{eq:normdiff}), 
the measured reference data
was slightly shifted by
\mbox{(-0.5, 2.0, 0.15)} 
pixels in 
($k_x$, $k_y$, $E$)
to correct for small differences in alignment.
To reduce noise, 
a Gaussian filter with a kernel (sigma) of
(\SI{0.005}{\per\angstrom}, \SI{0.005}{\per\angstrom}, 0.2 energy channels)
in
($k_x$, $k_y$, $E$)
was applied,
then both datasets were binned in groups of
(8, 8) 
pixels in 
($k_x$, $k_y$).
To select only the region of statistically relevant data, 
voxels with less than 80 counts in 
$Y_\text{SPP}$
or 
$Y_\text{photon}$
were ignored. 

\subsection*{Data Availability}
All calculated and experimental raw data from which the material in this work was evaluated is available on \url{doi:10.5281/zenodo.5102762}

\begin{acknowledgement} 
This work is funded by the Deutsche Forschungsgemeinschaft (DFG, German Research Foundation) - 
TRR 173 - 268565370 (Projects A02 and B11) 
and project RE1141/14-2. 
We thank the Nano Structuring Center (NSC) at TU Kaiserslautern for sample preparation.
Some simulations were executed on the high-performance cluster “Elwetritsch”
through the projects TopNano and Mulan at TU Kaiserslautern, which is a part of the “Alliance 
of High Performance Computing Rheinland-Pfalz”. We kindly acknowledge the support of 
Regionales Hochschulrechenzentrum Kaiserslautern.
E.P. acknowledges support from the Max Planck Graduate Center with the Johannes Gutenberg University Mainz and TU Kaiserslautern through a PhD fellowship.
M.H. would like to thank Benito Arnoldi and Andrew Price for help with 3D design and animation in Blender.
\end{acknowledgement} 

\begin{suppinfo} 
Real space PEEM images,
plots of the individual source term components,
derivation of the observed SPP velocity,
raw electron spectra and difference spectrum,
individual momentum microscopy data of the photon and SPP contributions,
additional momentum microscopy experimental results with mitigation of the secondary electron artifact,
movies of an animated spatiotemporal separation scheme and the time series of real space PEEM data.
\end{suppinfo} 

\bibliography{references.bib}

\providecommand{\latin}[1]{#1}
\makeatletter
\providecommand{\doi}
  {\begingroup\let\do\@makeother\dospecials
  \catcode`\{=1 \catcode`\}=2 \doi@aux}
\providecommand{\doi@aux}[1]{\endgroup\texttt{#1}}
\makeatother
\providecommand*\mcitethebibliography{\thebibliography}
\csname @ifundefined\endcsname{endmcitethebibliography}
  {\let\endmcitethebibliography\endthebibliography}{}
\begin{mcitethebibliography}{66}
\providecommand*\natexlab[1]{#1}
\providecommand*\mciteSetBstSublistMode[1]{}
\providecommand*\mciteSetBstMaxWidthForm[2]{}
\providecommand*\mciteBstWouldAddEndPuncttrue
  {\def\EndOfBibitem{\unskip.}}
\providecommand*\mciteBstWouldAddEndPunctfalse
  {\let\EndOfBibitem\relax}
\providecommand*\mciteSetBstMidEndSepPunct[3]{}
\providecommand*\mciteSetBstSublistLabelBeginEnd[3]{}
\providecommand*\EndOfBibitem{}
\mciteSetBstSublistMode{f}
\mciteSetBstMaxWidthForm{subitem}{(\alph{mcitesubitemcount})}
\mciteSetBstSublistLabelBeginEnd
  {\mcitemaxwidthsubitemform\space}
  {\relax}
  {\relax}

\bibitem[Brongersma \latin{et~al.}(2015)Brongersma, Halas, and
  Nordlander]{Brongersma_2015_NatureNanotechnology}
Brongersma,~M.~L.; Halas,~N.~J.; Nordlander,~P. Plasmon-induced hot carrier
  science and technology. \emph{Nature Nanotechnology} \textbf{2015},
  \emph{10}, 25--34\relax
\mciteBstWouldAddEndPuncttrue
\mciteSetBstMidEndSepPunct{\mcitedefaultmidpunct}
{\mcitedefaultendpunct}{\mcitedefaultseppunct}\relax
\EndOfBibitem
\bibitem[Linic \latin{et~al.}(2011)Linic, Christopher, and
  Ingram]{Linic_2011_NatureMaterials}
Linic,~S.; Christopher,~P.; Ingram,~D.~B. Plasmonic-metal nanostructures for
  efficient conversion of solar to chemical energy. \emph{Nature Materials}
  \textbf{2011}, \emph{10}, 911--921\relax
\mciteBstWouldAddEndPuncttrue
\mciteSetBstMidEndSepPunct{\mcitedefaultmidpunct}
{\mcitedefaultendpunct}{\mcitedefaultseppunct}\relax
\EndOfBibitem
\bibitem[Atwater and Polman(2010)Atwater, and
  Polman]{Atwater_2010_NatureMaterials}
Atwater,~H.~A.; Polman,~A. Plasmonics for improved photovoltaic devices.
  \emph{Nature Materials} \textbf{2010}, \emph{9}, 205--213\relax
\mciteBstWouldAddEndPuncttrue
\mciteSetBstMidEndSepPunct{\mcitedefaultmidpunct}
{\mcitedefaultendpunct}{\mcitedefaultseppunct}\relax
\EndOfBibitem
\bibitem[Aizpurua \latin{et~al.}(2019)Aizpurua, Baletto, Baumberg, Christopher,
  de~Nijs, Deshpande, Fernandez, Fabris, Freakley, Gawinkowski, Govorov, Halas,
  Hernandez, Jankiewicz, Khurgin, Kuisma, Kumar, Lischner, Liu, Marini, Maurer,
  Mueller, Parente, Park, Reich, Sivan, Tagliabue, Torrente-Murciano,
  Thangamuthu, Xiao, and Zayats]{faradaydiscussions-theoryhotelectrons}
Aizpurua,~J.; Baletto,~F.; Baumberg,~J.; Christopher,~P.; de~Nijs,~B.;
  Deshpande,~P.; Fernandez,~Y.~D.; Fabris,~L.; Freakley,~S.; Gawinkowski,~S.;
  Govorov,~A.; Halas,~N.; Hernandez,~R.; Jankiewicz,~B.; Khurgin,~J.;
  Kuisma,~M.; Kumar,~P.~V.; Lischner,~J.; Liu,~J.; Marini,~A. \latin{et~al.}
  Theory of hot electrons: general discussion. \emph{Faraday Discussions}
  \textbf{2019}, \emph{214}, 245--281\relax
\mciteBstWouldAddEndPuncttrue
\mciteSetBstMidEndSepPunct{\mcitedefaultmidpunct}
{\mcitedefaultendpunct}{\mcitedefaultseppunct}\relax
\EndOfBibitem
\bibitem[Baumberg(2019)]{Baumberg_2019_FaradayDiscussions}
Baumberg,~J.~J. Hot electron science in plasmonics and catalysis: what we argue
  about. \emph{Faraday Discussions} \textbf{2019}, \emph{214}, 501--511\relax
\mciteBstWouldAddEndPuncttrue
\mciteSetBstMidEndSepPunct{\mcitedefaultmidpunct}
{\mcitedefaultendpunct}{\mcitedefaultseppunct}\relax
\EndOfBibitem
\bibitem[Manjavacas \latin{et~al.}(2014)Manjavacas, Liu, Kulkarni, and
  Nordlander]{Manjavacas_2014_ACSNano}
Manjavacas,~A.; Liu,~J.~G.; Kulkarni,~V.; Nordlander,~P. Plasmon-Induced Hot
  Carriers in Metallic Nanoparticles. \emph{ACS Nano} \textbf{2014}, \emph{8},
  7630--7638\relax
\mciteBstWouldAddEndPuncttrue
\mciteSetBstMidEndSepPunct{\mcitedefaultmidpunct}
{\mcitedefaultendpunct}{\mcitedefaultseppunct}\relax
\EndOfBibitem
\bibitem[Sundararaman \latin{et~al.}(2014)Sundararaman, Narang, Jermyn,
  Goddard~III, and Atwater]{Sundararaman2014}
Sundararaman,~R.; Narang,~P.; Jermyn,~A.~S.; Goddard~III,~W.~A.; Atwater,~H.~A.
  Theoretical predictions for hot-carrier generation from surface plasmon
  decay. \emph{Nature Communications} \textbf{2014}, \emph{5}, 5788\relax
\mciteBstWouldAddEndPuncttrue
\mciteSetBstMidEndSepPunct{\mcitedefaultmidpunct}
{\mcitedefaultendpunct}{\mcitedefaultseppunct}\relax
\EndOfBibitem
\bibitem[Zhang and Govorov(2014)Zhang, and Govorov]{Zhang2014}
Zhang,~H.; Govorov,~A.~O. Optical Generation of Hot Plasmonic Carriers in Metal
  Nanocrystals: The Effects of Shape and Field Enhancement. \emph{The Journal
  of Physical Chemistry C} \textbf{2014}, \emph{118}, 7606--7614\relax
\mciteBstWouldAddEndPuncttrue
\mciteSetBstMidEndSepPunct{\mcitedefaultmidpunct}
{\mcitedefaultendpunct}{\mcitedefaultseppunct}\relax
\EndOfBibitem
\bibitem[Bernardi \latin{et~al.}(2015)Bernardi, Mustafa, Neaton, and
  Louie]{Bernardi_2015_Naturecommunications}
Bernardi,~M.; Mustafa,~J.; Neaton,~J.~B.; Louie,~S.~G. Theory and computation
  of hot carriers generated by surface plasmon polaritons in noble metals.
  \emph{Nature communications} \textbf{2015}, \emph{6}\relax
\mciteBstWouldAddEndPuncttrue
\mciteSetBstMidEndSepPunct{\mcitedefaultmidpunct}
{\mcitedefaultendpunct}{\mcitedefaultseppunct}\relax
\EndOfBibitem
\bibitem[Brown \latin{et~al.}(2016)Brown, Sundararaman, Narang, William
  A.~Goddard, and Atwater]{brown2016acsnano}
Brown,~A.~M.; Sundararaman,~R.; Narang,~P.; William A.~Goddard,~I.;
  Atwater,~H.~A. Nonradiative Plasmon Decay and Hot Carrier Dynamics: Effects
  of Phonons, Surfaces, and Geometry. \emph{ACS Nano} \textbf{2016}, \emph{10},
  957--966\relax
\mciteBstWouldAddEndPuncttrue
\mciteSetBstMidEndSepPunct{\mcitedefaultmidpunct}
{\mcitedefaultendpunct}{\mcitedefaultseppunct}\relax
\EndOfBibitem
\bibitem[Saavedra \latin{et~al.}(2016)Saavedra, Asenjo-Garcia, and
  Garc{\'i}a~de Abajo]{Saavedra_2016_ACSPhotonics}
Saavedra,~J. R.~M.; Asenjo-Garcia,~A.; Garc{\'i}a~de Abajo,~F.~J. Hot-Electron
  Dynamics and Thermalization in Small Metallic Nanoparticles. \emph{ACS
  Photonics} \textbf{2016}, \emph{3}, 1637--1646\relax
\mciteBstWouldAddEndPuncttrue
\mciteSetBstMidEndSepPunct{\mcitedefaultmidpunct}
{\mcitedefaultendpunct}{\mcitedefaultseppunct}\relax
\EndOfBibitem
\bibitem[Sykes \latin{et~al.}(2017)Sykes, Stewart, Akselrod, Kong, Wang,
  Gosztola, Martinson, Rosenmann, Mikkelsen, Govorov, and
  Wiederrecht]{Sykes2017}
Sykes,~M.~E.; Stewart,~J.~W.; Akselrod,~G.~M.; Kong,~X.-T.; Wang,~Z.;
  Gosztola,~D.~J.; Martinson,~A. B.~F.; Rosenmann,~D.; Mikkelsen,~M.~H.;
  Govorov,~A.~O.; Wiederrecht,~G.~P. Enhanced generation and anisotropic
  Coulomb scattering of hot electrons in an ultra-broadband plasmonic nanopatch
  metasurface. \emph{Nature Communications} \textbf{2017}, \emph{8}\relax
\mciteBstWouldAddEndPuncttrue
\mciteSetBstMidEndSepPunct{\mcitedefaultmidpunct}
{\mcitedefaultendpunct}{\mcitedefaultseppunct}\relax
\EndOfBibitem
\bibitem[Khurgin(2019)]{Khurgin2019faradaydiscussions}
Khurgin,~J.~B. Hot carriers generated by plasmons: where are they generated and
  where do they go from there? \emph{Faraday Discussions} \textbf{2019},
  \emph{214}, 35--58\relax
\mciteBstWouldAddEndPuncttrue
\mciteSetBstMidEndSepPunct{\mcitedefaultmidpunct}
{\mcitedefaultendpunct}{\mcitedefaultseppunct}\relax
\EndOfBibitem
\bibitem[Aguirregabiria \latin{et~al.}(2019)Aguirregabiria, Marinica, Ludwig,
  Brida, Leitenstorfer, Aizpurua, and
  Borisov]{Aguirregabiria_2019_FaradayDiscussions}
Aguirregabiria,~G.; Marinica,~D.-C.; Ludwig,~M.; Brida,~D.; Leitenstorfer,~A.;
  Aizpurua,~J.; Borisov,~A.~G. Dynamics of electron-emission currents in
  plasmonic gaps induced by strong fields. \emph{Faraday Discussions}
  \textbf{2019}, \emph{214}, 147--157\relax
\mciteBstWouldAddEndPuncttrue
\mciteSetBstMidEndSepPunct{\mcitedefaultmidpunct}
{\mcitedefaultendpunct}{\mcitedefaultseppunct}\relax
\EndOfBibitem
\bibitem[Do \latin{et~al.}(2021)Do, Jun, Mahfoud, Lin, and
  Bosman]{Do_2021_Nanoscale}
Do,~H. T.~B.; Jun,~D.~W.; Mahfoud,~Z.; Lin,~W.; Bosman,~M. Electron dynamics in
  plasmons. \emph{Nanoscale} \textbf{2021}, \emph{13}, 2801--2810\relax
\mciteBstWouldAddEndPuncttrue
\mciteSetBstMidEndSepPunct{\mcitedefaultmidpunct}
{\mcitedefaultendpunct}{\mcitedefaultseppunct}\relax
\EndOfBibitem
\bibitem[Benhayoun \latin{et~al.}(2021)Benhayoun, Terekhin, Ivanov, Rethfeld,
  and Garcia]{Benhayoun_2021_AppliedSurfaceScience}
Benhayoun,~O.; Terekhin,~P.~N.; Ivanov,~D.~S.; Rethfeld,~B.; Garcia,~M.~E.
  Theory for heating of metals assisted by Surface Plasmon Polaritons.
  \emph{Applied Surface Science} \textbf{2021}, 150427\relax
\mciteBstWouldAddEndPuncttrue
\mciteSetBstMidEndSepPunct{\mcitedefaultmidpunct}
{\mcitedefaultendpunct}{\mcitedefaultseppunct}\relax
\EndOfBibitem
\bibitem[Khurgin \latin{et~al.}(2021)Khurgin, Petrov, Eich, and
  Uskov]{Khurgin_2021_ACSPhotonics}
Khurgin,~J.~B.; Petrov,~A.; Eich,~M.; Uskov,~A.~V. Direct Plasmonic Excitation
  of the Hybridized Surface States in Metal Nanoparticles. \emph{{ACS}
  Photonics} \textbf{2021}, \relax
\mciteBstWouldAddEndPunctfalse
\mciteSetBstMidEndSepPunct{\mcitedefaultmidpunct}
{}{\mcitedefaultseppunct}\relax
\EndOfBibitem
\bibitem[Not()]{Note-1}
In this frequency range, the real part of the dielectric function passes zero
  (\textit{epsilon near zero}).\relax
\mciteBstWouldAddEndPunctfalse
\mciteSetBstMidEndSepPunct{\mcitedefaultmidpunct}
{}{\mcitedefaultseppunct}\relax
\EndOfBibitem
\bibitem[Hopfield(1965)]{Hopfield_1965_PhysicalReview}
Hopfield,~J.~J. Effect of Electron-Electron Interactions on Photoemission in
  Simple Metals. \emph{Physical Review} \textbf{1965}, \emph{139},
  A419--A424\relax
\mciteBstWouldAddEndPuncttrue
\mciteSetBstMidEndSepPunct{\mcitedefaultmidpunct}
{\mcitedefaultendpunct}{\mcitedefaultseppunct}\relax
\EndOfBibitem
\bibitem[Barman \latin{et~al.}(2004)Barman, Biswas, and
  Horn]{Barman_2004_SurfaceScience}
Barman,~S.~R.; Biswas,~C.; Horn,~K. Collective excitations on silver surfaces
  studied by photoyield. \emph{Surface Science} \textbf{2004}, \emph{566-568},
  538--543\relax
\mciteBstWouldAddEndPuncttrue
\mciteSetBstMidEndSepPunct{\mcitedefaultmidpunct}
{\mcitedefaultendpunct}{\mcitedefaultseppunct}\relax
\EndOfBibitem
\bibitem[Barman \latin{et~al.}(2004)Barman, Biswas, and
  Horn]{Barman_2004_PhysicalReviewB}
Barman,~S.; Biswas,~C.; Horn,~K. Electronic excitations on silver surfaces.
  \emph{Physical Review B} \textbf{2004}, \emph{69}, 045413\relax
\mciteBstWouldAddEndPuncttrue
\mciteSetBstMidEndSepPunct{\mcitedefaultmidpunct}
{\mcitedefaultendpunct}{\mcitedefaultseppunct}\relax
\EndOfBibitem
\bibitem[Reutzel \latin{et~al.}(2019)Reutzel, Li, Gumhalter, and
  Petek]{Reutzel_2019_PhysicalReviewLetters}
Reutzel,~M.; Li,~A.; Gumhalter,~B.; Petek,~H. Nonlinear Plasmonic Photoelectron
  Response of Ag(111). \emph{Physical Review Letters} \textbf{2019},
  \emph{123}, 017404\relax
\mciteBstWouldAddEndPuncttrue
\mciteSetBstMidEndSepPunct{\mcitedefaultmidpunct}
{\mcitedefaultendpunct}{\mcitedefaultseppunct}\relax
\EndOfBibitem
\bibitem[Li \latin{et~al.}(2021)Li, Reutzel, Wang, Novko, Gumhalter, and
  Petek]{Li_2021_ACSPhotonics}
Li,~A.; Reutzel,~M.; Wang,~Z.; Novko,~D.; Gumhalter,~B.; Petek,~H. Plasmonic
  Photoemission from Single-Crystalline Silver. \emph{{ACS} Photonics}
  \textbf{2021}, \emph{8}, 247--258\relax
\mciteBstWouldAddEndPuncttrue
\mciteSetBstMidEndSepPunct{\mcitedefaultmidpunct}
{\mcitedefaultendpunct}{\mcitedefaultseppunct}\relax
\EndOfBibitem
\bibitem[Novko \latin{et~al.}(2021)Novko, Despoja, Reutzel, Li, Petek, and
  Gumhalter]{Novko_2021_PhysicalReviewB}
Novko,~D.; Despoja,~V.; Reutzel,~M.; Li,~A.; Petek,~H.; Gumhalter,~B.
  Plasmonically assisted channels of photoemission from metals. \emph{Physical
  Review B} \textbf{2021}, \emph{103}, 205401\relax
\mciteBstWouldAddEndPuncttrue
\mciteSetBstMidEndSepPunct{\mcitedefaultmidpunct}
{\mcitedefaultendpunct}{\mcitedefaultseppunct}\relax
\EndOfBibitem
\bibitem[Kale and Christopher(2015)Kale, and Christopher]{Kale_2015_Science}
Kale,~M.~J.; Christopher,~P. Plasmons at the interface. \emph{Science}
  \textbf{2015}, \emph{349}, 587--588\relax
\mciteBstWouldAddEndPuncttrue
\mciteSetBstMidEndSepPunct{\mcitedefaultmidpunct}
{\mcitedefaultendpunct}{\mcitedefaultseppunct}\relax
\EndOfBibitem
\bibitem[Tan \latin{et~al.}(2017)Tan, Argondizzo, Ren, Liu, Zhao, and
  Petek]{Tan_2017_NaturePhotonics}
Tan,~S.; Argondizzo,~A.; Ren,~J.; Liu,~L.; Zhao,~J.; Petek,~H. Plasmonic
  coupling at a metal/semiconductor interface. \emph{Nature Photonics}
  \textbf{2017}, \emph{11}, 806--812\relax
\mciteBstWouldAddEndPuncttrue
\mciteSetBstMidEndSepPunct{\mcitedefaultmidpunct}
{\mcitedefaultendpunct}{\mcitedefaultseppunct}\relax
\EndOfBibitem
\bibitem[Tan \latin{et~al.}(2018)Tan, Dai, Zhang, Liu, Zhao, and
  Petek]{Tan_2018_PhysicalReviewLetters}
Tan,~S.; Dai,~Y.; Zhang,~S.; Liu,~L.; Zhao,~J.; Petek,~H. Coherent Electron
  Transfer at the Ag/Graphite Heterojunction Interface. \emph{Physical Review
  Letters} \textbf{2018}, \emph{120}, 126801\relax
\mciteBstWouldAddEndPuncttrue
\mciteSetBstMidEndSepPunct{\mcitedefaultmidpunct}
{\mcitedefaultendpunct}{\mcitedefaultseppunct}\relax
\EndOfBibitem
\bibitem[Shibuta \latin{et~al.}(2021)Shibuta, Yamamoto, Ohta, Inoue, Mizoguchi,
  Nakaya, Eguchi, and Nakajima]{Shibuta_2021_ACSNano}
Shibuta,~M.; Yamamoto,~K.; Ohta,~T.; Inoue,~T.; Mizoguchi,~K.; Nakaya,~M.;
  Eguchi,~T.; Nakajima,~A. Confined Hot Electron Relaxation at the Molecular
  Heterointerface of the Size-Selected Plasmonic Noble Metal Nanocluster and
  Layered C60. \emph{{ACS} Nano} \textbf{2021}, \emph{15}, 1199--1209\relax
\mciteBstWouldAddEndPuncttrue
\mciteSetBstMidEndSepPunct{\mcitedefaultmidpunct}
{\mcitedefaultendpunct}{\mcitedefaultseppunct}\relax
\EndOfBibitem
\bibitem[Link and El-Sayed(1999)Link, and El-Sayed]{link1999}
Link,~S.; El-Sayed,~M.~A. Spectral Properties and Relaxation Dynamics of
  Surface Plasmon Electronic Oscillations in Gold and Silver Nanodots and
  Nanorods. \emph{The Journal of Physical Chemistry B} \textbf{1999},
  \emph{103}, 8410--8426\relax
\mciteBstWouldAddEndPuncttrue
\mciteSetBstMidEndSepPunct{\mcitedefaultmidpunct}
{\mcitedefaultendpunct}{\mcitedefaultseppunct}\relax
\EndOfBibitem
\bibitem[Foerster \latin{et~al.}(2017)Foerster, Joplin, Kaefer, Celiksoy, Link,
  and S\"{o}nnichsen]{foerster2017}
Foerster,~B.; Joplin,~A.; Kaefer,~K.; Celiksoy,~S.; Link,~S.;
  S\"{o}nnichsen,~C. Chemical Interface Damping Depends on Electrons Reaching
  the Surface. \emph{ACS Nano} \textbf{2017}, \emph{11}, 2886--2893\relax
\mciteBstWouldAddEndPuncttrue
\mciteSetBstMidEndSepPunct{\mcitedefaultmidpunct}
{\mcitedefaultendpunct}{\mcitedefaultseppunct}\relax
\EndOfBibitem
\bibitem[Harutyunyan \latin{et~al.}(2015)Harutyunyan, Martinson, Rosenmann,
  Khorashad, Besteiro, Govorov, and Wiederrecht]{Harutyunyan2015}
Harutyunyan,~H.; Martinson,~A. B.~F.; Rosenmann,~D.; Khorashad,~L.~K.;
  Besteiro,~L.~V.; Govorov,~A.~O.; Wiederrecht,~G.~P. Anomalous ultrafast
  dynamics of hot plasmonic electrons in nanostructures with hot spots.
  \emph{Nature Nanotechnology} \textbf{2015}, \emph{10}, 770--774\relax
\mciteBstWouldAddEndPuncttrue
\mciteSetBstMidEndSepPunct{\mcitedefaultmidpunct}
{\mcitedefaultendpunct}{\mcitedefaultseppunct}\relax
\EndOfBibitem
\bibitem[M{\'{e}}jard \latin{et~al.}(2016)M{\'{e}}jard, Verdy, Petit,
  Bouhelier, Cluzel, and Demichel]{Mjard2016}
M{\'{e}}jard,~R.; Verdy,~A.; Petit,~M.; Bouhelier,~A.; Cluzel,~B.; Demichel,~O.
  Energy-Resolved Hot-Carrier Relaxation Dynamics in Monocrystalline Plasmonic
  Nanoantennas. \emph{{ACS} Photonics} \textbf{2016}, \emph{3},
  1482--1488\relax
\mciteBstWouldAddEndPuncttrue
\mciteSetBstMidEndSepPunct{\mcitedefaultmidpunct}
{\mcitedefaultendpunct}{\mcitedefaultseppunct}\relax
\EndOfBibitem
\bibitem[Heilpern \latin{et~al.}(2018)Heilpern, Manjare, Govorov, Wiederrecht,
  Gray, and Harutyunyan]{Heilpern_2018_NatureCommunications}
Heilpern,~T.; Manjare,~M.; Govorov,~A.~O.; Wiederrecht,~G.~P.; Gray,~S.~K.;
  Harutyunyan,~H. Determination of hot carrier energy distributions from
  inversion of ultrafast pump-probe reflectivity measurements. \emph{Nature
  Communications} \textbf{2018}, \emph{9}\relax
\mciteBstWouldAddEndPuncttrue
\mciteSetBstMidEndSepPunct{\mcitedefaultmidpunct}
{\mcitedefaultendpunct}{\mcitedefaultseppunct}\relax
\EndOfBibitem
\bibitem[Fujimoto \latin{et~al.}(1984)Fujimoto, Liu, Ippen, and
  Bloembergen]{Fujimoto1984}
Fujimoto,~J.~G.; Liu,~J.~M.; Ippen,~E.~P.; Bloembergen,~N. Femtosecond Laser
  Interaction with Metallic Tungsten and Nonequilibrium Electron and Lattice
  Temperatures. \emph{Physical Review Letters} \textbf{1984}, \emph{53},
  1837--1840\relax
\mciteBstWouldAddEndPuncttrue
\mciteSetBstMidEndSepPunct{\mcitedefaultmidpunct}
{\mcitedefaultendpunct}{\mcitedefaultseppunct}\relax
\EndOfBibitem
\bibitem[Schoenlein \latin{et~al.}(1988)Schoenlein, Fujimoto, Eesley, and
  Capehart]{Schoenlein1988}
Schoenlein,~R.~W.; Fujimoto,~J.~G.; Eesley,~G.~L.; Capehart,~T.~W. Femtosecond
  Studies of Image-Potential Dynamics in Metals. \emph{Physical Review Letters}
  \textbf{1988}, \emph{61}, 2596--2599\relax
\mciteBstWouldAddEndPuncttrue
\mciteSetBstMidEndSepPunct{\mcitedefaultmidpunct}
{\mcitedefaultendpunct}{\mcitedefaultseppunct}\relax
\EndOfBibitem
\bibitem[Schmuttenmaer \latin{et~al.}(1994)Schmuttenmaer, Aeschlimann,
  Elsayed-Ali, Miller, Mantell, Cao, and Gao]{Schmuttenmaer1994}
Schmuttenmaer,~C.~A.; Aeschlimann,~M.; Elsayed-Ali,~H.~E.; Miller,~R. J.~D.;
  Mantell,~D.~A.; Cao,~J.; Gao,~Y. Time-resolved two-photon photoemission from
  Cu(100): Energy dependence of electron relaxation. \emph{Physical Review B}
  \textbf{1994}, \emph{50}, 8957--8960\relax
\mciteBstWouldAddEndPuncttrue
\mciteSetBstMidEndSepPunct{\mcitedefaultmidpunct}
{\mcitedefaultendpunct}{\mcitedefaultseppunct}\relax
\EndOfBibitem
\bibitem[Petek and Ogawa(1997)Petek, and Ogawa]{petek1997}
Petek,~H.; Ogawa,~S. Femtosecond time-resolved two-photon photoemission studies
  of electron dynamics in metals. \emph{Progress in Surface Science}
  \textbf{1997}, \emph{56}, 239 -- 310\relax
\mciteBstWouldAddEndPuncttrue
\mciteSetBstMidEndSepPunct{\mcitedefaultmidpunct}
{\mcitedefaultendpunct}{\mcitedefaultseppunct}\relax
\EndOfBibitem
\bibitem[Bauer \latin{et~al.}(2015)Bauer, Marienfeld, and
  Aeschlimann]{Bauer2015319}
Bauer,~M.; Marienfeld,~A.; Aeschlimann,~M. Hot electron lifetimes in metals
  probed by time-resolved two-photon photoemission. \emph{Progress in Surface
  Science} \textbf{2015}, \emph{90}, 319 -- 376\relax
\mciteBstWouldAddEndPuncttrue
\mciteSetBstMidEndSepPunct{\mcitedefaultmidpunct}
{\mcitedefaultendpunct}{\mcitedefaultseppunct}\relax
\EndOfBibitem
\bibitem[Aeschlimann \latin{et~al.}(1996)Aeschlimann, Bauer, and
  Pawlik]{ma1996}
Aeschlimann,~M.; Bauer,~M.; Pawlik,~S. Competing nonradiative channels for hot
  electron induced surface photochemistry. \emph{Chemical Physics}
  \textbf{1996}, \emph{205}, 127 -- 141\relax
\mciteBstWouldAddEndPuncttrue
\mciteSetBstMidEndSepPunct{\mcitedefaultmidpunct}
{\mcitedefaultendpunct}{\mcitedefaultseppunct}\relax
\EndOfBibitem
\bibitem[Schmidt \latin{et~al.}(2002)Schmidt, Bauer, Wiemann, Porath, Scharte,
  Andreyev, Sch{\"{o}}nhense, and Aeschlimann]{Schmidt_2002_ApplPhysB}
Schmidt,~O.; Bauer,~M.; Wiemann,~C.; Porath,~R.; Scharte,~M.; Andreyev,~O.;
  Sch{\"{o}}nhense,~G.; Aeschlimann,~M. Time-resolved two photon photoemission
  electron microscopy. \emph{Appl Phys B} \textbf{2002}, \emph{74},
  223--227\relax
\mciteBstWouldAddEndPuncttrue
\mciteSetBstMidEndSepPunct{\mcitedefaultmidpunct}
{\mcitedefaultendpunct}{\mcitedefaultseppunct}\relax
\EndOfBibitem
\bibitem[Cinchetti \latin{et~al.}(2005)Cinchetti, Gloskovskii, Nepjiko,
  Sch{\"o}nhense, Rochholz, and Kreiter]{Cinchetti2005peem}
Cinchetti,~M.; Gloskovskii,~A.; Nepjiko,~S.~A.; Sch{\"o}nhense,~G.;
  Rochholz,~H.; Kreiter,~M. Photoemission Electron Microscopy as a Tool for the
  Investigation of Optical Near Fields. \emph{Physical Review Letters}
  \textbf{2005}, \emph{95}, 047601\relax
\mciteBstWouldAddEndPuncttrue
\mciteSetBstMidEndSepPunct{\mcitedefaultmidpunct}
{\mcitedefaultendpunct}{\mcitedefaultseppunct}\relax
\EndOfBibitem
\bibitem[Bayer \latin{et~al.}(2008)Bayer, Wiemann, Gaier, Bauer, and
  Aeschlimann]{bayer2008time}
Bayer,~D.; Wiemann,~C.; Gaier,~O.; Bauer,~M.; Aeschlimann,~M. Time-resolved
  2PPE and time-resolved PEEM as a probe of LSP's in silver nanoparticles.
  \emph{Journal of Nanomaterials} \textbf{2008}, \emph{2008}\relax
\mciteBstWouldAddEndPuncttrue
\mciteSetBstMidEndSepPunct{\mcitedefaultmidpunct}
{\mcitedefaultendpunct}{\mcitedefaultseppunct}\relax
\EndOfBibitem
\bibitem[Lemke \latin{et~al.}(2014)Lemke, Lei{\ss}ner, Evlyukhin, Radke, Klick,
  Fiutowski, Kjelstrup-Hansen, Rubahn, Chichkov, Reinhardt, and
  Bauer]{Lemke_2014_NanoLetters}
Lemke,~C.; Lei{\ss}ner,~T.; Evlyukhin,~A.; Radke,~J.~W.; Klick,~A.;
  Fiutowski,~J.; Kjelstrup-Hansen,~J.; Rubahn,~H.-G.; Chichkov,~B.~N.;
  Reinhardt,~C.; Bauer,~M. The Interplay between Localized and Propagating
  Plasmonic Excitations Tracked in Space and Time. \emph{Nano Letters}
  \textbf{2014}, \emph{14}, 2431--2435\relax
\mciteBstWouldAddEndPuncttrue
\mciteSetBstMidEndSepPunct{\mcitedefaultmidpunct}
{\mcitedefaultendpunct}{\mcitedefaultseppunct}\relax
\EndOfBibitem
\bibitem[Kahl \latin{et~al.}(2014)Kahl, Wall, Witt, Schneider, Bayer, Fischer,
  Melchior, Horn-von Hoegen, Aeschlimann, and Meyer~zu Heringdorf]{Kahl2014}
Kahl,~P.; Wall,~S.; Witt,~C.; Schneider,~C.; Bayer,~D.; Fischer,~A.;
  Melchior,~P.; Horn-von Hoegen,~M.; Aeschlimann,~M.; Meyer~zu
  Heringdorf,~F.-J. Normal-Incidence Photoemission Electron Microscopy
  (NI-PEEM) for Imaging Surface Plasmon Polaritons. \emph{Plasmonics}
  \textbf{2014}, \emph{9}, 1401\relax
\mciteBstWouldAddEndPuncttrue
\mciteSetBstMidEndSepPunct{\mcitedefaultmidpunct}
{\mcitedefaultendpunct}{\mcitedefaultseppunct}\relax
\EndOfBibitem
\bibitem[Kahl \latin{et~al.}(2017)Kahl, Podbiel, Schneider, Makris, Sindermann,
  Witt, Kilbane, Horn-von Hoegen, Aeschlimann, and Meyer~zu
  Heringdorf]{Kahl_2017_Plasmonics}
Kahl,~P.; Podbiel,~D.; Schneider,~C.; Makris,~A.; Sindermann,~S.; Witt,~C.;
  Kilbane,~D.; Horn-von Hoegen,~M.; Aeschlimann,~M.; Meyer~zu Heringdorf,~F.
  Direct Observation of Surface Plasmon Polariton Propagation and Interference
  by Time-Resolved Imaging in Normal-Incidence Two Photon Photoemission
  Microscopy. \emph{Plasmonics} \textbf{2017}, \emph{13}, 239--246\relax
\mciteBstWouldAddEndPuncttrue
\mciteSetBstMidEndSepPunct{\mcitedefaultmidpunct}
{\mcitedefaultendpunct}{\mcitedefaultseppunct}\relax
\EndOfBibitem
\bibitem[Scharte \latin{et~al.}(2001)Scharte, Porath, Ohms, Aeschlimann, Krenn,
  Ditlbacher, Aussenegg, and Liebsch]{Scharte2001}
Scharte,~M.; Porath,~R.; Ohms,~T.; Aeschlimann,~M.; Krenn,~J.; Ditlbacher,~H.;
  Aussenegg,~F.; Liebsch,~A. Do Mie plasmons have a longer lifetime on
  resonance than off resonance? \emph{Applied Physics B} \textbf{2001},
  \emph{73}, 305--310\relax
\mciteBstWouldAddEndPuncttrue
\mciteSetBstMidEndSepPunct{\mcitedefaultmidpunct}
{\mcitedefaultendpunct}{\mcitedefaultseppunct}\relax
\EndOfBibitem
\bibitem[Merschdorf \latin{et~al.}(2004)Merschdorf, Kennerknecht, and
  Pfeiffer]{Merschdorf_2004_Phys.Rev.B}
Merschdorf,~M.; Kennerknecht,~C.; Pfeiffer,~W. Collective and single-particle
  dynamics in time-resolved two-photon photoemission. \emph{Physical Review B}
  \textbf{2004}, \emph{70}, 193401\relax
\mciteBstWouldAddEndPuncttrue
\mciteSetBstMidEndSepPunct{\mcitedefaultmidpunct}
{\mcitedefaultendpunct}{\mcitedefaultseppunct}\relax
\EndOfBibitem
\bibitem[Podbiel \latin{et~al.}(2017)Podbiel, Kahl, Makris, Frank, Sindermann,
  Davis, Giessen, von Hoegen, and zu~Heringdorf]{Podbiel2017}
Podbiel,~D.; Kahl,~P.; Makris,~A.; Frank,~B.; Sindermann,~S.; Davis,~T.~J.;
  Giessen,~H.; von Hoegen,~M.~H.; zu~Heringdorf,~F.-J.~M. Imaging the Nonlinear
  Plasmoemission Dynamics of Electrons from Strong Plasmonic Fields. \emph{Nano
  Letters} \textbf{2017}, \emph{17}, 6569--6574\relax
\mciteBstWouldAddEndPuncttrue
\mciteSetBstMidEndSepPunct{\mcitedefaultmidpunct}
{\mcitedefaultendpunct}{\mcitedefaultseppunct}\relax
\EndOfBibitem
\bibitem[Lehr \latin{et~al.}(2017)Lehr, Foerster, Schmitt, Kr{\"{u}}ger,
  S\"{o}nnichsen, Sch{\"{o}}nhense, and Elmers]{Lehr_2017_NanoLetters}
Lehr,~M.; Foerster,~B.; Schmitt,~M.; Kr{\"{u}}ger,~K.; S\"{o}nnichsen,~C.;
  Sch{\"{o}}nhense,~G.; Elmers,~H.-J. Momentum Distribution of Electrons
  Emitted from Resonantly Excited Individual Gold Nanorods. \emph{Nano Letters}
  \textbf{2017}, \emph{17}, 6606--6612\relax
\mciteBstWouldAddEndPuncttrue
\mciteSetBstMidEndSepPunct{\mcitedefaultmidpunct}
{\mcitedefaultendpunct}{\mcitedefaultseppunct}\relax
\EndOfBibitem
\bibitem[Lehr \latin{et~al.}(2019)Lehr, Bley, Vogel, Rethfeld, Sch\"{o}nhense,
  and Elmers]{Lehr_2019_TheJournalofPhysicalChemistryC}
Lehr,~M.; Bley,~K.; Vogel,~N.; Rethfeld,~B.; Sch\"{o}nhense,~G.; Elmers,~H.-J.
  Evidence of Spatially Inhomogeneous Electron Temperature in a Resonantly
  Excited Array of Bow-Tie Nanoantennas. \emph{The Journal of Physical
  Chemistry C} \textbf{2019}, \emph{123}, 12429--12436\relax
\mciteBstWouldAddEndPuncttrue
\mciteSetBstMidEndSepPunct{\mcitedefaultmidpunct}
{\mcitedefaultendpunct}{\mcitedefaultseppunct}\relax
\EndOfBibitem
\bibitem[Pettine \latin{et~al.}(2021)Pettine, Meyer, Medeghini, Murphy, and
  Nesbitt]{Pettine_2021_ACSnano}
Pettine,~J.; Meyer,~S.~M.; Medeghini,~F.; Murphy,~C.~J.; Nesbitt,~D.~J.
  Controlling the Spatial and Momentum Distributions of Plasmonic Carriers:
  Volume vs Surface Effects. \emph{{ACS} Nano} \textbf{2021}, \emph{15},
  1566--1578\relax
\mciteBstWouldAddEndPuncttrue
\mciteSetBstMidEndSepPunct{\mcitedefaultmidpunct}
{\mcitedefaultendpunct}{\mcitedefaultseppunct}\relax
\EndOfBibitem
\bibitem[Spektor \latin{et~al.}(2019)Spektor, Kilbane, Mahro, Hartelt, Prinz,
  Aeschlimann, and Orenstein]{Spektor_2019_PhysicalReviewX}
Spektor,~G.; Kilbane,~D.; Mahro,~A.; Hartelt,~M.; Prinz,~E.; Aeschlimann,~M.;
  Orenstein,~M. Mixing the Light Spin with Plasmon Orbit by Nonlinear
  Light-Matter Interaction in Gold. \emph{Physical Review X} \textbf{2019},
  \emph{9}, 021031\relax
\mciteBstWouldAddEndPuncttrue
\mciteSetBstMidEndSepPunct{\mcitedefaultmidpunct}
{\mcitedefaultendpunct}{\mcitedefaultseppunct}\relax
\EndOfBibitem
\bibitem[Not()]{Note-2}
A corresponding pulse propagating in negative $x$-direction is also launched
  from the slit, but it is outside of the detection area.\relax
\mciteBstWouldAddEndPunctfalse
\mciteSetBstMidEndSepPunct{\mcitedefaultmidpunct}
{}{\mcitedefaultseppunct}\relax
\EndOfBibitem
\bibitem[Not()]{Note-3}
The exact scaling depends on dephasing of the coherences along the excitation
  pathway.\relax
\mciteBstWouldAddEndPunctfalse
\mciteSetBstMidEndSepPunct{\mcitedefaultmidpunct}
{}{\mcitedefaultseppunct}\relax
\EndOfBibitem
\bibitem[Terekhin \latin{et~al.}(2020)Terekhin, Benhayoun, Weber, Ivanov,
  Garcia, and Rethfeld]{Terekhin_2020_AppliedSurfaceScience}
Terekhin,~P.~N.; Benhayoun,~O.; Weber,~S.~T.; Ivanov,~D.~S.; Garcia,~M.~E.;
  Rethfeld,~B. Influence of surface plasmon polaritons on laser energy
  absorption and structuring of surfaces. \emph{Applied Surface Science}
  \textbf{2020}, \emph{512}, 144420\relax
\mciteBstWouldAddEndPuncttrue
\mciteSetBstMidEndSepPunct{\mcitedefaultmidpunct}
{\mcitedefaultendpunct}{\mcitedefaultseppunct}\relax
\EndOfBibitem
\bibitem[Joly \latin{et~al.}(2018)Joly, El-Khoury, and
  Hess]{Joly_2018_TheJournalofPhysicalChemistryC}
Joly,~A.~G.; El-Khoury,~P.~Z.; Hess,~W.~P. Spatiotemporal Imaging of Surface
  Plasmons Using Two-Color Photoemission Electron Microscopy. \emph{The Journal
  of Physical Chemistry C} \textbf{2018}, \emph{122}, 20981--20988\relax
\mciteBstWouldAddEndPuncttrue
\mciteSetBstMidEndSepPunct{\mcitedefaultmidpunct}
{\mcitedefaultendpunct}{\mcitedefaultseppunct}\relax
\EndOfBibitem
\bibitem[Kotsugi \latin{et~al.}(2003)Kotsugi, Kuch, Offi, Chelaru, and
  Kirschner]{Kotsugi_2003_ReviewofScientificInstruments}
Kotsugi,~M.; Kuch,~W.; Offi,~F.; Chelaru,~L.~I.; Kirschner,~J.
  Microspectroscopic two-dimensional Fermi surface mapping using a
  photoelectron emission microscope. \emph{Review of Scientific Instruments}
  \textbf{2003}, \emph{74}, 2754--2758\relax
\mciteBstWouldAddEndPuncttrue
\mciteSetBstMidEndSepPunct{\mcitedefaultmidpunct}
{\mcitedefaultendpunct}{\mcitedefaultseppunct}\relax
\EndOfBibitem
\bibitem[Tusche \latin{et~al.}(2016)Tusche, Goslawski, Kutnyakhov, Ellguth,
  Medjanik, Elmers, Chernov, Wallauer, Engel, Jankowiak, and
  Sch\"{o}nhense]{Tusche_2016_AppliedPhysicsLetters}
Tusche,~C.; Goslawski,~P.; Kutnyakhov,~D.; Ellguth,~M.; Medjanik,~K.;
  Elmers,~H.~J.; Chernov,~S.; Wallauer,~R.; Engel,~D.; Jankowiak,~A.;
  Sch\"{o}nhense,~G. Multi-{MHz} time-of-flight electronic bandstructure
  imaging of graphene on Ir(111). \emph{Applied Physics Letters} \textbf{2016},
  \emph{108}, 261602\relax
\mciteBstWouldAddEndPuncttrue
\mciteSetBstMidEndSepPunct{\mcitedefaultmidpunct}
{\mcitedefaultendpunct}{\mcitedefaultseppunct}\relax
\EndOfBibitem
\bibitem[Haag \latin{et~al.}(2019)Haag, Eul, Thielen, Haag, Stadtm{\"{u}}ller,
  and Aeschlimann]{Haag_2019_ReviewofScientificInstruments}
Haag,~F.; Eul,~T.; Thielen,~P.; Haag,~N.; Stadtm{\"{u}}ller,~B.;
  Aeschlimann,~M. Time-resolved two-photon momentum microscopy - A new approach
  to study hot carrier lifetimes in momentum space. \emph{Review of Scientific
  Instruments} \textbf{2019}, \emph{90}, 103104\relax
\mciteBstWouldAddEndPuncttrue
\mciteSetBstMidEndSepPunct{\mcitedefaultmidpunct}
{\mcitedefaultendpunct}{\mcitedefaultseppunct}\relax
\EndOfBibitem
\bibitem[Maklar \latin{et~al.}(2020)Maklar, Dong, Beaulieu, Pincelli, Dendzik,
  Windsor, Xian, Wolf, Ernstorfer, and
  Rettig]{Maklar_2020_ReviewofScientificInstruments}
Maklar,~J.; Dong,~S.; Beaulieu,~S.; Pincelli,~T.; Dendzik,~M.; Windsor,~Y.~W.;
  Xian,~R.~P.; Wolf,~M.; Ernstorfer,~R.; Rettig,~L. A quantitative comparison
  of time-of-flight momentum microscopes and hemispherical analyzers for time-
  and angle-resolved photoemission spectroscopy experiments. \emph{Review of
  Scientific Instruments} \textbf{2020}, \emph{91}, 123112\relax
\mciteBstWouldAddEndPuncttrue
\mciteSetBstMidEndSepPunct{\mcitedefaultmidpunct}
{\mcitedefaultendpunct}{\mcitedefaultseppunct}\relax
\EndOfBibitem
\bibitem[Wu \latin{et~al.}(2015)Wu, Chen, McBride, and Lian]{Wu_2015_Science}
Wu,~K.; Chen,~J.; McBride,~J.~R.; Lian,~T. Efficient hot-electron transfer by a
  plasmon-induced interfacial charge-transfer transition. \emph{Science}
  \textbf{2015}, \emph{349}, 632--635\relax
\mciteBstWouldAddEndPuncttrue
\mciteSetBstMidEndSepPunct{\mcitedefaultmidpunct}
{\mcitedefaultendpunct}{\mcitedefaultseppunct}\relax
\EndOfBibitem
\bibitem[Therrien \latin{et~al.}(2019)Therrien, Kale, Yuan, Zhang, Halas, and
  Christopher]{Therrien_2019_FaradayDiscussions}
Therrien,~A.~J.; Kale,~M.~J.; Yuan,~L.; Zhang,~C.; Halas,~N.~J.;
  Christopher,~P. Impact of chemical interface damping on surface plasmon
  dephasing. \emph{Faraday Discussions} \textbf{2019}, \emph{214}, 59--72\relax
\mciteBstWouldAddEndPuncttrue
\mciteSetBstMidEndSepPunct{\mcitedefaultmidpunct}
{\mcitedefaultendpunct}{\mcitedefaultseppunct}\relax
\EndOfBibitem
\bibitem[Foerster \latin{et~al.}(2020)Foerster, Hartelt, Collins, Aeschlimann,
  Link, and S{\"{o}}nnichsen]{Foerster_2020_NanoLetters}
Foerster,~B.; Hartelt,~M.; Collins,~S. S.~E.; Aeschlimann,~M.; Link,~S.;
  S{\"{o}}nnichsen,~C. Interfacial States Cause Equal Decay of Plasmons and Hot
  Electrons at Gold{\textendash}Metal Oxide Interfaces. \emph{Nano Letters}
  \textbf{2020}, \emph{20}, 3338--3343\relax
\mciteBstWouldAddEndPuncttrue
\mciteSetBstMidEndSepPunct{\mcitedefaultmidpunct}
{\mcitedefaultendpunct}{\mcitedefaultseppunct}\relax
\EndOfBibitem
\bibitem[Olmon \latin{et~al.}(2012)Olmon, Slovick, Johnson, Shelton, Oh,
  Boreman, and Raschke]{Olmon_2012_Phys.Rev.B}
Olmon,~R.~L.; Slovick,~B.; Johnson,~T.~W.; Shelton,~D.; Oh,~S.-H.;
  Boreman,~G.~D.; Raschke,~M.~B. Optical dielectric function of gold.
  \emph{Physical Review B} \textbf{2012}, \emph{86}, 235147\relax
\mciteBstWouldAddEndPuncttrue
\mciteSetBstMidEndSepPunct{\mcitedefaultmidpunct}
{\mcitedefaultendpunct}{\mcitedefaultseppunct}\relax
\EndOfBibitem
\bibitem[Oelsner \latin{et~al.}(2010)Oelsner, Rohmer, Schneider, Bayer,
  Sch{\"o}nhense, and Aeschlimann]{Oelsner2010317}
Oelsner,~A.; Rohmer,~M.; Schneider,~C.; Bayer,~D.; Sch{\"o}nhense,~G.;
  Aeschlimann,~M. Time- and energy resolved photoemission electron
  microscopy-imaging of photoelectron time-of-flight analysis by means of
  pulsed excitations. \emph{Journal of Electron Spectroscopy and Related
  Phenomena} \textbf{2010}, \emph{178 - 179}, 317 -- 330\relax
\mciteBstWouldAddEndPuncttrue
\mciteSetBstMidEndSepPunct{\mcitedefaultmidpunct}
{\mcitedefaultendpunct}{\mcitedefaultseppunct}\relax
\EndOfBibitem
\end{mcitethebibliography}


\begin{thebibliography}{2}%
\makeatletter
\providecommand \@ifxundefined [1]{%
 \@ifx{#1\undefined}
}%
\providecommand \@ifnum [1]{%
 \ifnum #1\expandafter \@firstoftwo
 \else \expandafter \@secondoftwo
 \fi
}%
\providecommand \@ifx [1]{%
 \ifx #1\expandafter \@firstoftwo
 \else \expandafter \@secondoftwo
 \fi
}%
\providecommand \natexlab [1]{#1}%
\providecommand \enquote  [1]{``#1''}%
\providecommand \bibnamefont  [1]{#1}%
\providecommand \bibfnamefont [1]{#1}%
\providecommand \citenamefont [1]{#1}%
\providecommand \href@noop [0]{\@secondoftwo}%
\providecommand \href [0]{\begingroup \@sanitize@url \@href}%
\providecommand \@href[1]{\@@startlink{#1}\@@href}%
\providecommand \@@href[1]{\endgroup#1\@@endlink}%
\providecommand \@sanitize@url [0]{\catcode `\\12\catcode `\$12\catcode
  `\&12\catcode `\#12\catcode `\^12\catcode `\_12\catcode `\%12\relax}%
\providecommand \@@startlink[1]{}%
\providecommand \@@endlink[0]{}%
\providecommand \url  [0]{\begingroup\@sanitize@url \@url }%
\providecommand \@url [1]{\endgroup\@href {#1}{\urlprefix }}%
\providecommand \urlprefix  [0]{URL }%
\providecommand \Eprint [0]{\href }%
\providecommand \doibase [0]{http://dx.doi.org/}%
\providecommand \selectlanguage [0]{\@gobble}%
\providecommand \bibinfo  [0]{\@secondoftwo}%
\providecommand \bibfield  [0]{\@secondoftwo}%
\providecommand \translation [1]{[#1]}%
\providecommand \BibitemOpen [0]{}%
\providecommand \bibitemStop [0]{}%
\providecommand \bibitemNoStop [0]{.\EOS\space}%
\providecommand \EOS [0]{\spacefactor3000\relax}%
\providecommand \BibitemShut  [1]{\csname bibitem#1\endcsname}%
\let\auto@bib@innerbib\@empty
\bibitem [{\citenamefont {Maier}(2007)}]{maierplasmonics}%
  \BibitemOpen
  \bibfield  {author} {\bibinfo {author} {\bibfnamefont {S.~A.}\ \bibnamefont
  {Maier}},\ }\href {http://link.springer.com/book/10.1007%2F0-387-37825-1}
  {\emph {\bibinfo {title} {Plasmonics: fundamentals and applications:
  fundamentals and applications}}}\ (\bibinfo  {publisher} {Springer Science \&
  Business Media},\ \bibinfo {year} {2007})\BibitemShut {NoStop}%
\bibitem [{\citenamefont {Kubo}\ \emph {et~al.}(2007)\citenamefont {Kubo},
  \citenamefont {Pontius},\ and\ \citenamefont
  {Petek}}]{Kubo_2007_NanoLetters}%
  \BibitemOpen
  \bibfield  {author} {\bibinfo {author} {\bibfnamefont {A.}~\bibnamefont
  {Kubo}}, \bibinfo {author} {\bibfnamefont {N.}~\bibnamefont {Pontius}}, \
  and\ \bibinfo {author} {\bibfnamefont {H.}~\bibnamefont {Petek}},\ }\href
  {\doibase 10.1021/nl0627846} {\bibfield  {journal} {\bibinfo  {journal} {Nano
  Letters}\ }\textbf {\bibinfo {volume} {7}},\ \bibinfo {pages} {470} (\bibinfo
  {year} {2007})}\BibitemShut {NoStop}%
\end{thebibliography}%


\providecommand{\latin}[1]{#1}
\makeatletter
\providecommand{\doi}
  {\begingroup\let\do\@makeother\dospecials
  \catcode`\{=1 \catcode`\}=2 \doi@aux}
\providecommand{\doi@aux}[1]{\endgroup\texttt{#1}}
\makeatother
\providecommand*\mcitethebibliography{\thebibliography}
\csname @ifundefined\endcsname{endmcitethebibliography}
  {\let\endmcitethebibliography\endthebibliography}{}
%

\end{document}


\newcommand{\titlestr}{Energy and Momentum Distribution of Surface Plasmon-induced Hot Carriers Isolated via Spatiotemporal Separation}

\author{Michael Hartelt} 
\email{hartelt@physik.uni-kl.de}

\author{Pavel N. Terekhin} 
\author{Tobias Eul} 
\author{Anna-Katharina Mahro} 
\author{Benjamin Frisch}
\author{Eva Prinz} 
\affiliation{Department of Physics and Research Center OPTIMAS, TU Kaiserslautern, Germany}


\author{Baerbel Rethfeld}
\affiliation{Department of Physics and Research Center OPTIMAS, TU Kaiserslautern, Germany}	

\author{Benjamin Stadtm\"uller} 
\affiliation{Department of Physics and Research Center OPTIMAS, TU Kaiserslautern, Germany}	

\author{Martin Aeschlimann} 
\affiliation{Department of Physics and Research Center OPTIMAS, TU Kaiserslautern, Germany}	

\title{\titlestr
		\\
		------
		\\
		Supplementary Information}

\date{\today}

\maketitle

\section{Real Space images of the sample}

\begin{figure}[h]
	\includegraphics[]{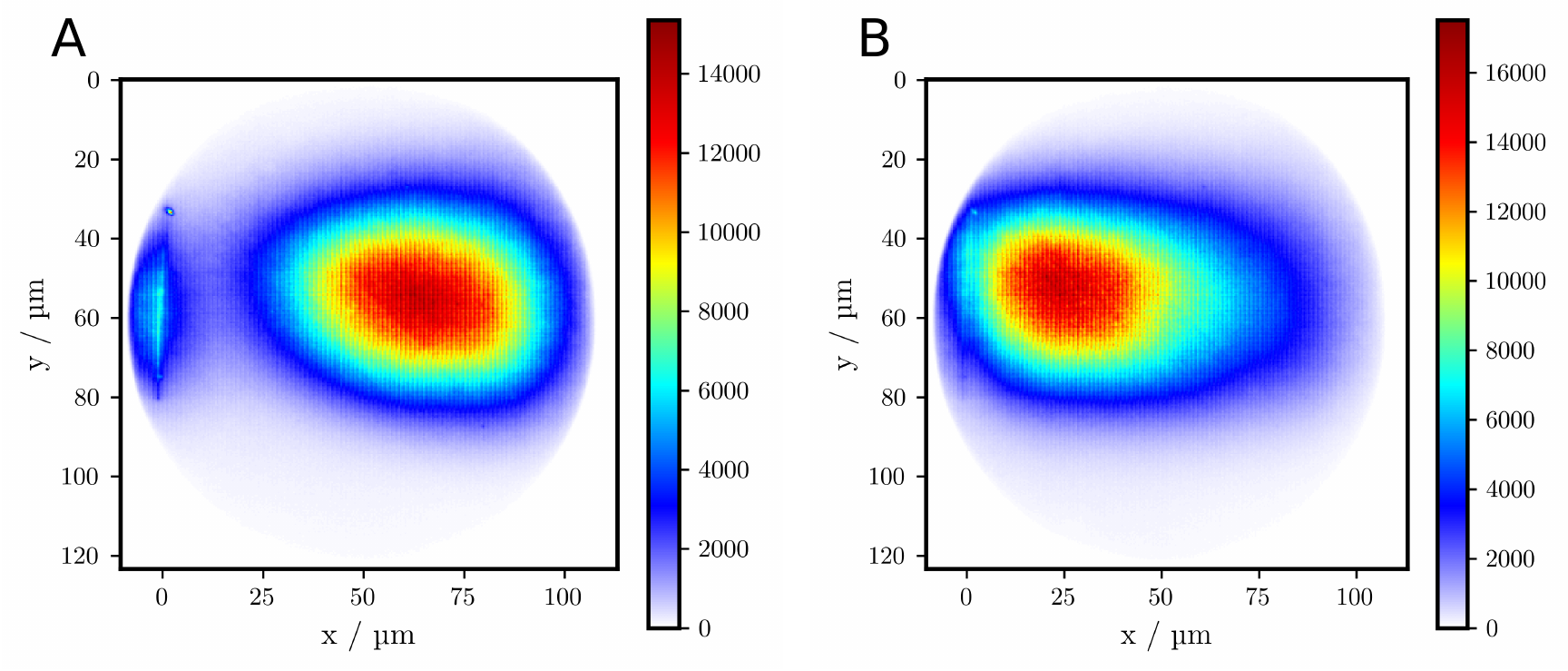}
	\caption{
		Real space PEEM images of the sample. 
		A) PEEM image of the static probe contribution $Y_{\text{probe}}$ to the photoemission signal, 
		taken from the full 4D dataset for $\Delta t = $\SIrange{-300}{-200}{\femto\second} (no temporal overlap), 
		integrated over all electron energies.
		Mainly visible is the spatial profile of the probe pulse on the right side 
		("Blue 2PPE")
		and the excitation slit 
		at $x = \SI{0}{\micro\meter}$ 
		illuminated by the pump pulse on the left side 
		("Red 3PPE").
		B) PEEM image, taken from the full 4D dataset for $\Delta t = $\SIrange{-10}{10}{\femto\second} (time zero), 
		integrated over all electron energies. 
		The enhanced yield 
		between the coupling edged and the probe pulse
		comes from electrons excited by the pump pulse 
		and photo-emitted by the probe pulse
		("Red-Blue 2PPE").
		}
	\label{fig:SI_realspace}
\end{figure}

\pagebreak

\section{Components of the Theoretical Source Term}

\begin{figure}[h]
	\includegraphics[width=\linewidth]{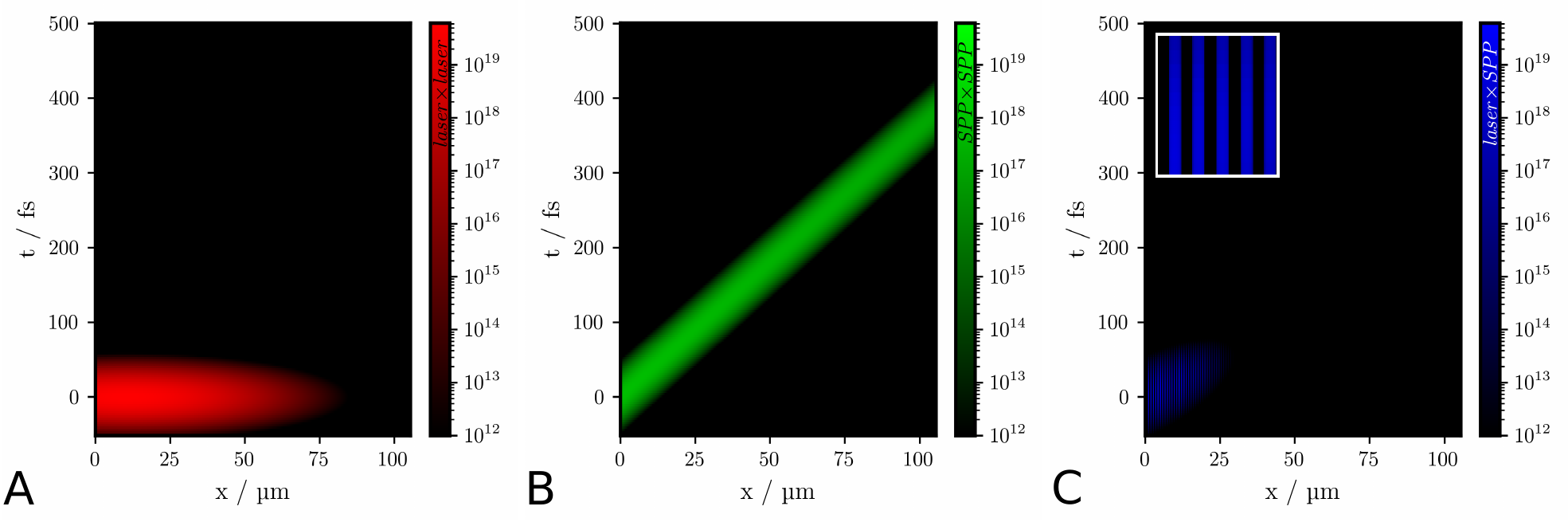}
	\caption{
		Calculation of the source term components 
		A) \textit{laser-laser}, 
		B) \textit{SPP-SPP}, and
		C) \textit{laser-SPP}.
		The time-dependent energy density 
		of the electromagnetic field contributions 
		at the surface is shown 
		for an excitation pulse with $\lambda = \SI{800}{\nano\meter}$ 
		and a pulse duration of \SI{23}{\femto\second}.
		The pulse maximum arrives at the sample surface at $t = \SI{0}{\femto\second}$.
		The colorbars are given in \si{\J \per \second \per \cubic \meter}.
		The inset in C) is a zoom of 
		$x =$ \SIrange{10}{15}{\micro\meter}
		and
		$t =$ \SIrange{15}{45}{\femto\second},
		showing the striped interference pattern.
		}
	\label{fig:SI_xtcomponents}
\end{figure}

\section{Calculation of the observed SPP velocity}

The real part of the SPP wave vector 
$k^\prime_\mathrm{SPP}$ 
from the solution of the wave equation at the interface 
\cite{maierplasmonics} 
is given by
\begin{equation}
	k^\prime_\mathrm{SPP} 
	= \Re\left(\frac{\omega}{c_0} \cdot \sqrt{\frac{\epsilon_\mathrm{m} \epsilon_\mathrm{d}}{\epsilon_\mathrm{m} + \epsilon_\mathrm{d}}}\right)
	\text{,}
\end{equation}
while the probe pulse counter-propagates with an in-plane wave vector of 
\begin{equation}
k_\parallel = - k_{\mathrm{probe}} \sin (\mathrm{AOI}) = - \frac{\omega}{c_0} \sin (\mathrm{AOI})
\text{,}
\end{equation}
the negative sign 
signifies the orientation of the projected wave vector in negative $x$-direction.

This 
results in the 
well-known effect for SPP waves observed in PEEM, 
\cite{Kubo_2007_NanoLetters} 
the so-called beating pattern
\begin{equation}
k_\mathrm{perceived} = k^\prime_\mathrm{SPP} - k_\parallel
\text{.}
\label{eq:SI_kperceived}
\end{equation}
We can link the perceived SPP group velocity 
$v_\text{perceived}$ 
and the actual SPP group velocity 
$v_\text{SPP}$ 
by taking the derivative 
over the frequency $\omega$ from 
equation \ref{eq:SI_kperceived}:
\begin{equation}
\frac{1}{v_\mathrm{perceived}}
= \frac{d k_\mathrm{perceived}}{d \omega}
= \frac{d k^\prime_\mathrm{SPP}}{d \omega} - \frac{d k_\parallel}{d \omega}
= \frac{1}{v_\mathrm{SPP}} + \frac{\sin (\mathrm{AOI})}{c_0}
\text{.}
\end{equation}
Solving for $v_\text{perceived}$ yields the relation used in the main manuscript:
\begin{equation}
v_\text{perceived}
= v_\text{SPP} \cdot \frac{1}{1 + \frac{v_\mathrm{SPP}}{c_0} \sin (\mathrm{AOI})}
\text{.}
\end{equation}

\section{Raw Spectra and Difference Spectrum}

\begin{figure}[h]
	\includegraphics[]{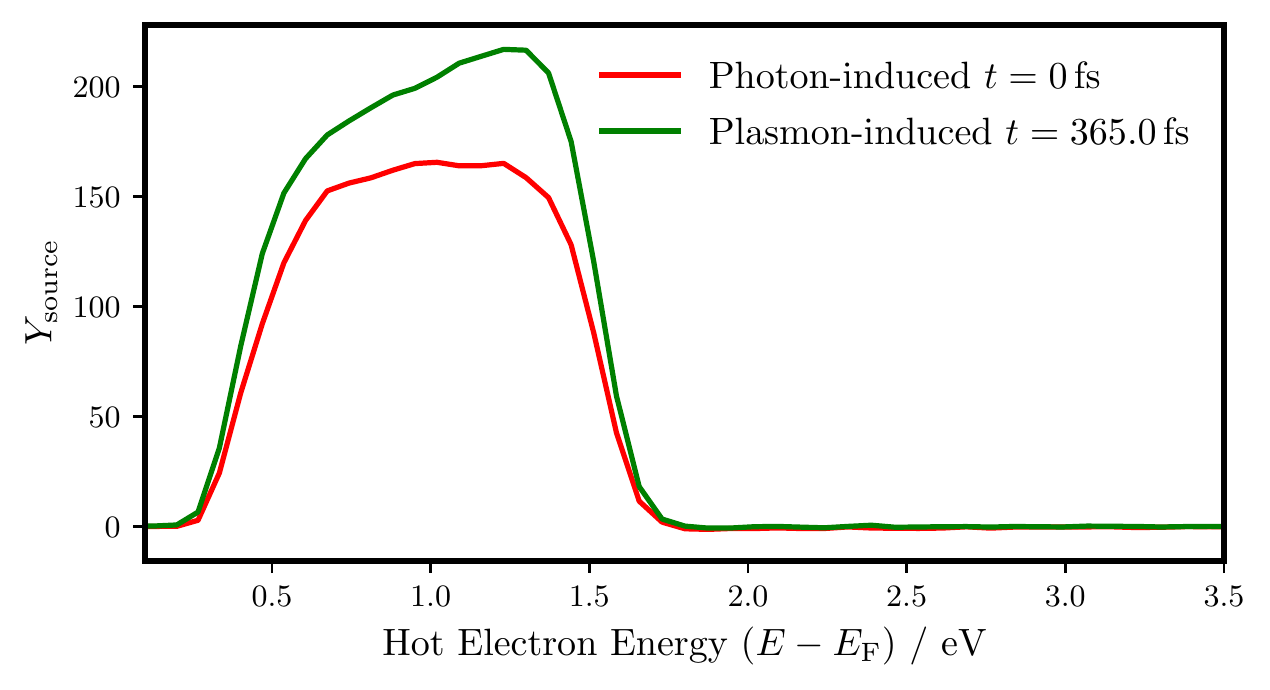}
	\caption{
		Raw Spectra from the $\Delta t$-Energy-plot in the main manuscript, 
		without normalization.
		The
		$Y_\text{source}$ values are given in counts.
		}
	\label{fig:SI_spectra}
\end{figure}

The 
photon-induced and plasmon-induced hot electron spectra 
from 
the $\Delta t$-Energy-plot in the main manuscript
are normalized 
to their work function edges
at the
low-energy cutoff 
for 
$E-E_\mathrm{F} = \SI{0.6}{\electronvolt}$.
For reference,
the raw spectra without normalization
are given in 
Figure \ref{fig:SI_spectra}.
The difference 
in absolute height
is
caused by the strength of the respective source field.

\begin{figure}[h]
	\includegraphics[]{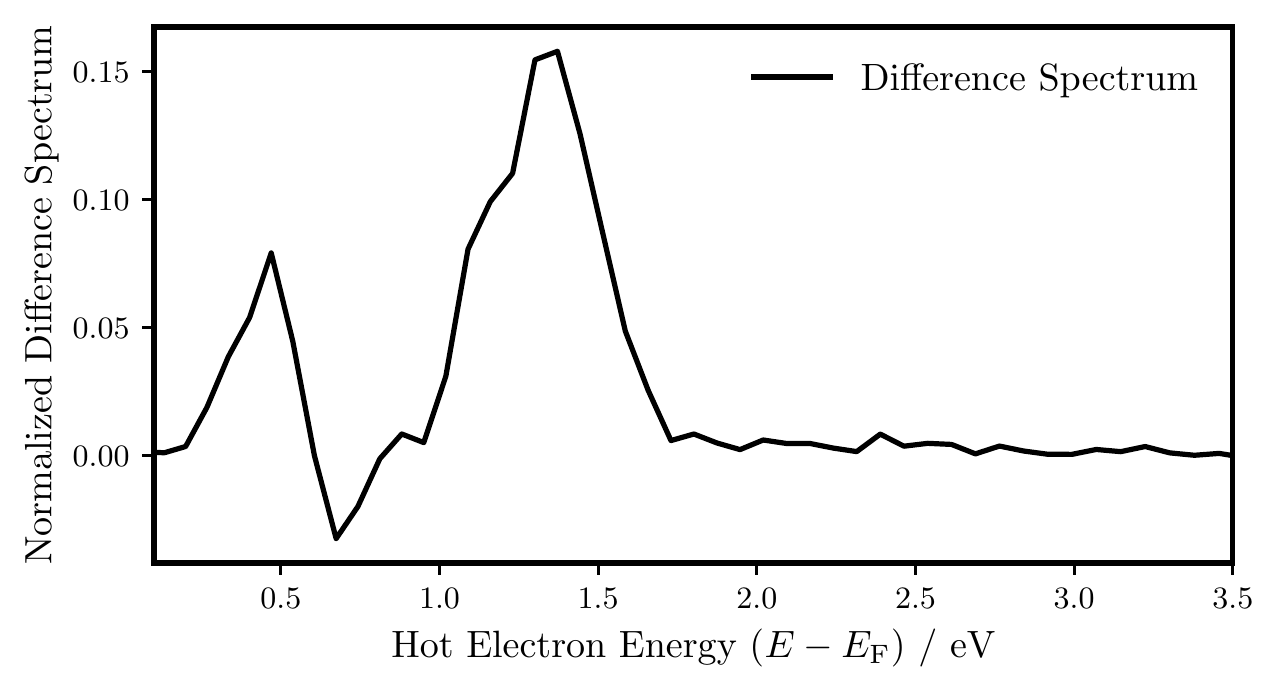}
	\caption{
		Difference Spectrum between
		$Y_{\text{source}}(\Delta t_\text{SPP})$
		and 
		$Y_{\text{source}}(\Delta t_\text{0})$ 
		as normalized in 
		the main manuscript.
		}
	\label{fig:SI_spectra_diff}
\end{figure}

The two normalized curves from 
the main manuscript
were subtracted for 
Figure \ref{fig:SI_spectra_diff}.
The main feature is 
the plasmonic excitation peak
at high energies.
According to the reasoning 
in the main manuscript, 
this would correspond to 
the electron spectrum 
of the \textit{plasmon} aspect of SPP.
The smaller peak 
in the low-energy cutoff region,
representing a slight difference 
in the shape of the work function edge,
is most likely caused
by the SPP propagation
through the finite real space integration area.
As discussed in
Section \ref{sec:SI_propagation_artifact}
this 
causes an artifact of increased detection of
secondary electrons.

\pagebreak

\section{Momentum Microscopy Data}
\begin{figure}[h]
	\includegraphics[width=\linewidth]{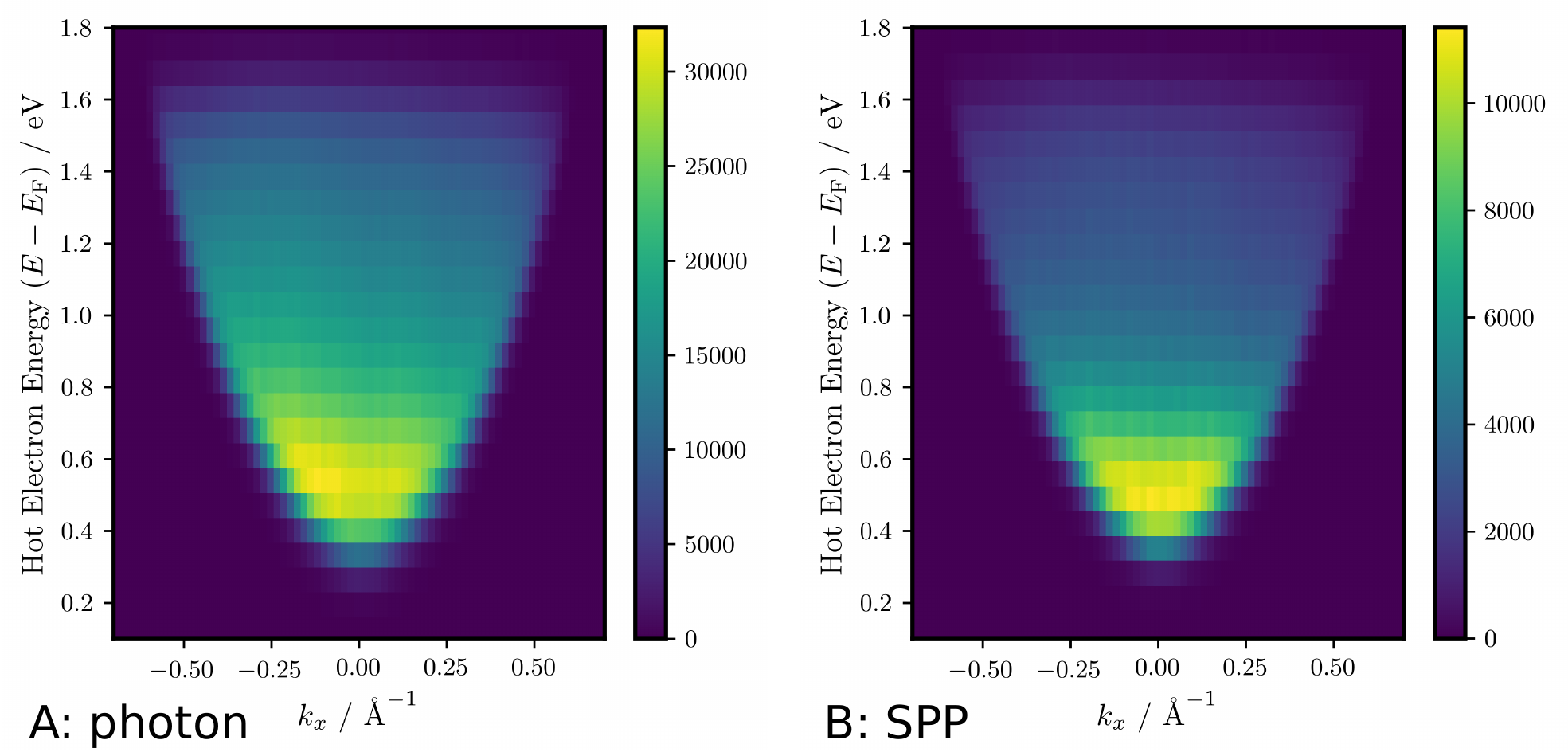}
	\caption{
		Momentum microscopy data
		A) $Y_{\text{photon}}(k_x,k_y,E)$
		and
		B) $Y_{\text{SPP}}(k_x,k_y,E)$
		along 
		the $E$-$k_x$-direction for $k_y = \SI{0}{\per \angstrom}$.
		}
	\label{fig:SI_kspace_e_kx}
\end{figure}

\begin{figure}[h]
	\includegraphics[width=\linewidth]{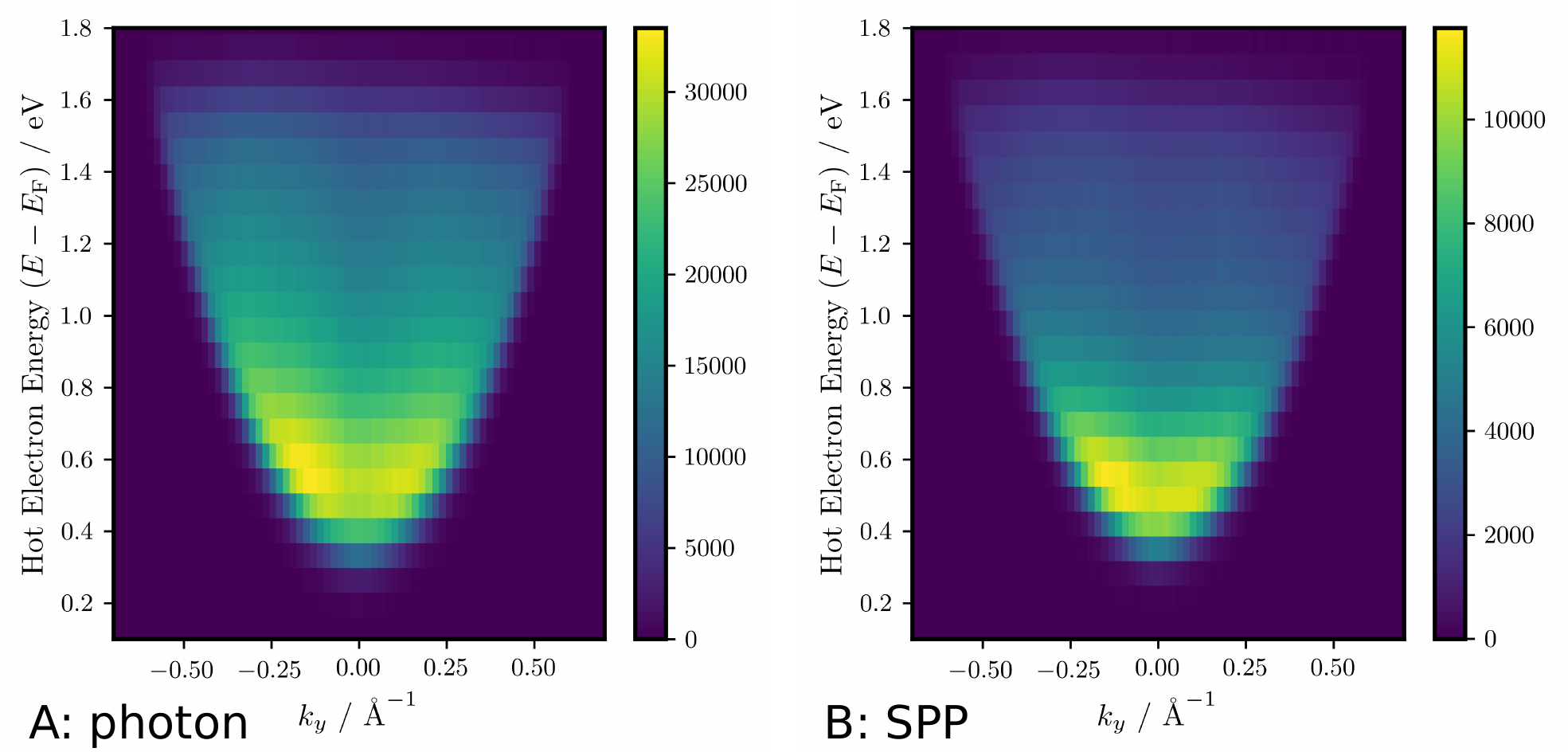}
	\caption{
		Momentum microscopy data
		A) $Y_{\text{photon}}(k_x,k_y,E)$
		and
		B) $Y_{\text{SPP}}(k_x,k_y,E)$
		along 
		the $E$-$k_y$-direction for $k_x = \SI{0}{\per \angstrom}$.
		}
	\label{fig:SI_kspace_e_ky}
\end{figure}

\begin{figure}[h]
	\includegraphics[width=\linewidth]{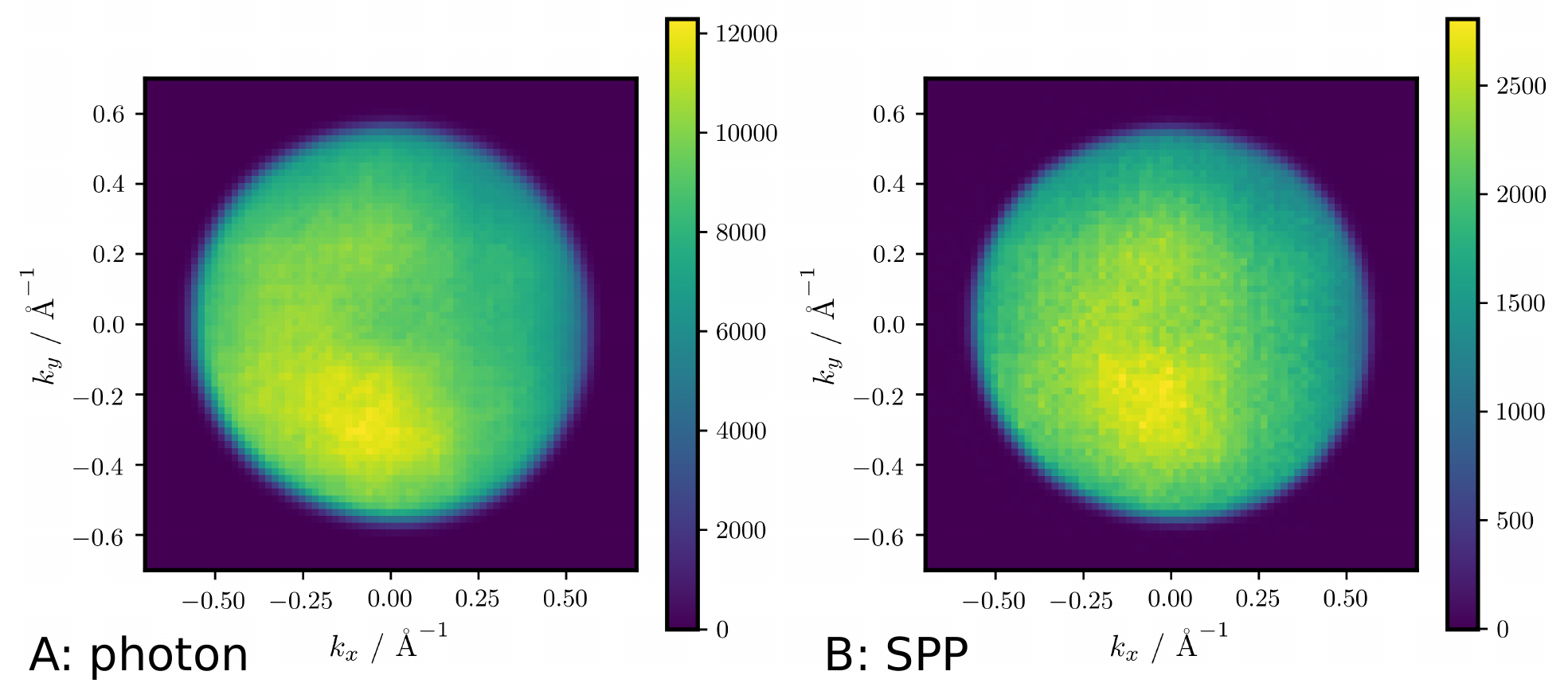}
	\caption{
		Momentum microscopy data
		A) $Y_{\text{photon}}(k_x,k_y,E)$
		and
		B) $Y_{\text{SPP}}(k_x,k_y,E)$
		along 
		the $k_x$-$k_y$-direction for $E - E_\mathrm{F} = $\SIrange{1.4}{1.5}{\electronvolt}.
		}
	\label{fig:SI_kspace_kx_ky}
\end{figure}

The momentum microscopy plots 
in figures \ref{fig:SI_kspace_e_kx}, \ref{fig:SI_kspace_e_ky}, and \ref{fig:SI_kspace_kx_ky}
were shifted, noise-filtered and binned
as described in under Data Evaluation 
in the Methods section in the main manuscript.

\section{Compensation of the SPP Propagation Artifact in the Momentum Distribution\label{sec:SI_propagation_artifact}}

\begin{figure}[h]
	\includegraphics[width=\linewidth]{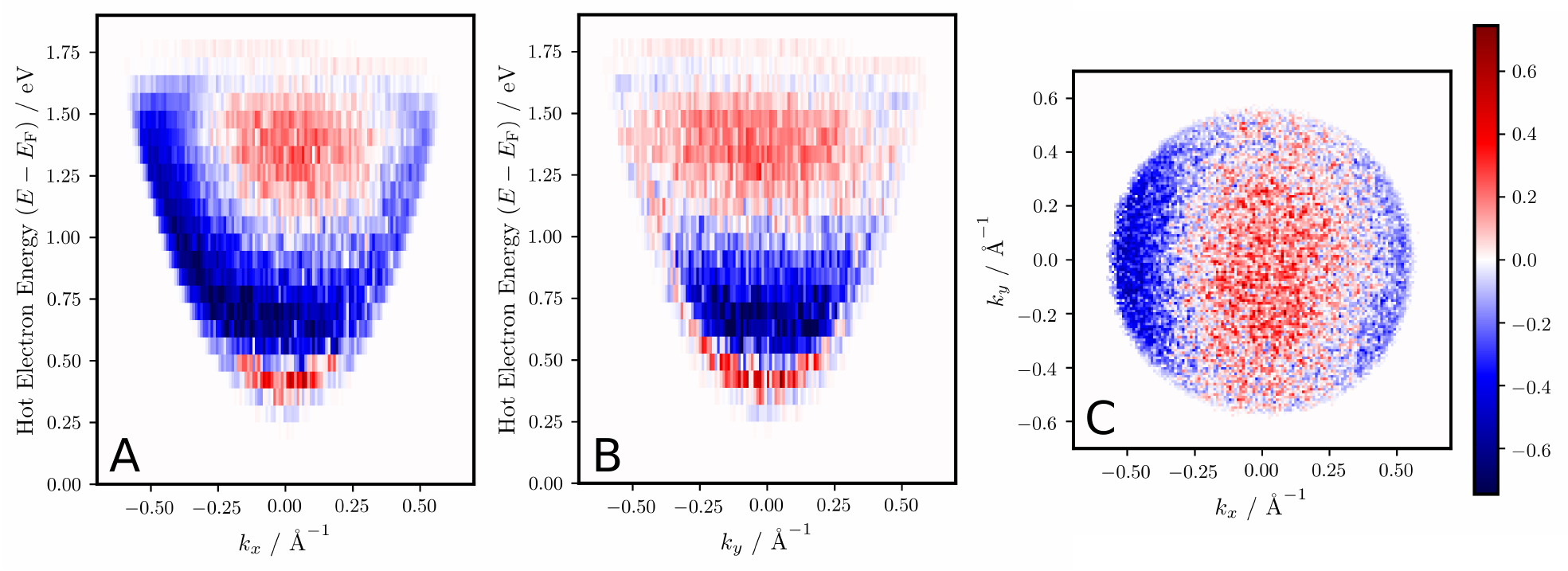}
	\caption{
		Momentum distribution of plasmon-induced hot electrons relative to photon-induced hot electrons,
		mitigating the apparent broadening effect by the propagation of the SPP pulse through the field-of-view.
		Plotted are cuts
		of the normalized difference values $\Delta Y(k_x, k_y, E)$
		along 
		A) the $E$-$k_x$-direction for $k_y = \SI{0}{\per \angstrom}$, 
		B) the $E$-$k_y$-direction for $k_x = \SI{0}{\per \angstrom}$, 
		and C) the $k_x$-$k_y$-direction for $E - E_\mathrm{F} = $\SIrange{1.4}{1.5}{\electronvolt}.
		}
	\label{fig:SI_kspacedata}
\end{figure}

As described in the main manuscript,
the apparent low-energy enhancement of 
plasmon-induced electrons 
in the $k$-space data
is caused by an apparent broadening 
of the SPP pulse duration
due to its propagation
through the detection area.
This is 
an observation artifact
caused by the experimental scheme and setup,
and not to be confused
with a real, dispersive pulse broadening.
(Due to the very linear SPP dispersion
over the used wavelength range,
the dispersive broadening
during the propagation time
is small.)
To mitigate this effect,
a measurement with 
reduced aperture size 
of \SI{\approx 18}{\micro\meter}
was performed,
recording the full time trace of 
$\Delta t =$ \SIrange{-334}{800}{\femto\second}.
The evaluation was carried out as described in the main manuscript,
but using integration ranges
instead of fixed values
for the respective delay times:
$\Delta t_\text{SPP} =$ \SIrange{345}{355}{\femto\second},
$\Delta t_\text{photon} =$ \SIrange{-80}{80}{\femto\second},
and
$\Delta t_\text{probe} =$ \SIrange{-334}{-167}{\femto\second}.
In this way, 
a prolonged time window 
for the photonic signal
compensates for the 
propagation time
of the SPP pulse
through the detection area.
To select only the region of statistically relevant data, 
voxels with 
less than 80 counts in 
$Y_\text{SPP}$
or 
less than 5 counts in 
$Y_\text{photon}$
were ignored. 

The result in 
Figure \ref{fig:SI_kspacedata}
shows that 
the larger integration range for the photonic signal
compensates for the propagation artifact
and the signal enhancement caused by secondary electrons
is mitigated.
The high-energy feature remains, 
but it
is deformed along the 
$k_y$-direction.
This deformation 
is caused by an electron-optical artifact
when the iris aperture 
is pulled to very small diameters.
It is then no longer symmetric,
but a hexagonal shape 
elongated along the 
$y$-direction.

\section{Movies}
\begin{itemize}
\item Animation of the Scheme of Spatiotemporal Separation
	\\
	\path{spatiotemporal_scheme_animated.mkv}
\item Movie of the Spatiotemporal Dynamics Plasmon-induced Hot Carriers.
	Shown is $Y_{\text{source}}(x,y,E,\Delta t)$
	for an electron energy of $E = \SI{1.5}{\electronvolt}$.
	The field-of-view and axes are as depicted in Figure \ref{fig:SI_realspace}.
	\\
	\path{timeseries_binned_fermi.avi}
\end{itemize}

\bibliography{references.bib}